\documentclass[11pt, a4paper]{article}
\usepackage[left=2cm, right=2cm, top=2cm, bottom=2cm]{geometry}
\usepackage[utf8]{inputenc}
\usepackage{amsmath}
\usepackage{amssymb}
\usepackage{multirow}
\usepackage{booktabs}
\usepackage{natbib}
\usepackage{hyperref}
\usepackage{amsthm}
\usepackage{verbatim}
\usepackage{graphicx}
\graphicspath{{fig/}}
\usepackage{subcaption}
\usepackage{enumitem}
\usepackage{textcomp}
\usepackage{adjustbox}
\usepackage{mathtools}
\usepackage{cancel}
\usepackage{placeins}
\usepackage{algorithmic,algorithm}
\usepackage{adjustbox}

\mathtoolsset{showonlyrefs}
\newcommand{\E}{\mathbb{E}}

\newcommand{\R}{\mathbb{R}}

\newcommand{\X}{\mathcal{X}}

\newcommand{\Y}{\mathcal{Y}}
\newcommand{\Z}{\mathcal{Z}}
\newcommand{\apprpost}{Q_\phi(\cdot|\ddobs_i)}
\newcommand{\apprpostl}{Q_{\phi,l}(\cdot|\ddobs_i)}

\newcommand{\ddz}{\mathbf{z}}
\newcommand{\Ddz}{\mathbf{Z}}
\newcommand{\ddobs}{\mathbf{y}}

\newcommand{\ddsim}{\mathbf{x}}
\newcommand{\Ddsim}{\mathbf{X}}

\newcommand{\Ddobs}{\mathbf{Y}}

\newcommand{\sumjk}{\sum_{\substack{j,k=1\\k\neq j}}^m}

\newcommand{\SE}{S_{\operatorname{E}}}

\newcommand{\sumij}{\sum_{\substack{j,k=1\\j\neq k}}^m}
\newcommand{\ddtheta}{{\boldsymbol{\theta}}}

\newcommand\independent{\protect\mathpalette{\protect\independenT}{\perp}}
\def\independenT#1#2{\mathrel{\rlap{$#1#2$}\mkern2mu{#1#2}}}

\DeclareMathOperator*{\argmin}{arg\,min}

\theoremstyle{definition}

\title{{Likelihood-Free Inference with Generative Neural Networks\\via Scoring Rule Minimization}} % Title: Limited to 180 characters in length. Use title capitalization.

\author{
	Lorenzo Pacchiardi$^1$\thanks{Corresponding author: lorenzo.pacchiardi@stats.ox.ac.uk.}, Ritabrata Dutta$^2 $\\
	{\em \small $^1$Department of Statistics, University of Oxford, UK}\\
	{\em \small $^2$Department of Statistics, University of Warwick, UK}\\
}
\date{}

\begin{document}
	
	\maketitle
	
	\begin{comment}
	\begin{abstract}
		Bayesian Likelihood-Free Inference methods yield approximate posteriors for simulator models with intractable likelihood.  
		Some methods approximate the posterior using normalizing flows, namely neural networks implementing invertible transformations. The density is then accessible and the neural network can be trained via 
		maximum likelihood from simulated parameter-observation pairs.  
		Here, we generalize this approach to Scoring Rules (SRs) minimization. SRs evaluate probability distributions with respect to an outcome; we employ them to assess how the approximate posterior for a simulated observation matches the corresponding parameter value.
		Some SRs are minimized when approximate and true posteriors coincide and can be estimated via samples from the approximate posterior. This allows relaxing the invertibility requirement and using more general neural networks.
	\end{abstract}
	\end{comment}

	\begin{abstract}
		Bayesian Likelihood-Free Inference methods yield posterior approximations for simulator models with intractable likelihood. Recently, many works trained neural networks to approximate either the intractable likelihood or the posterior directly. 
		Most proposals use normalizing flows, namely neural networks parametrizing invertible maps used to transform samples from an underlying base measure; the probability density of the transformed samples is then accessible and the normalizing flow can be trained via maximum likelihood on simulated parameter-observation pairs. A recent work \citep{ramesh2022gatsbi} approximated instead the posterior with generative networks, which drop the invertibility requirement and are thus a more flexible class of distributions scaling to high-dimensional and structured data. However, generative networks only allow sampling from the parametrized distribution; for this reason, \cite{ramesh2022gatsbi} follows the common solution of adversarial training, where the generative network plays a min-max game against a ``{critic}'' network.
%		objective is iteratively optimized
		 This procedure is unstable and can lead to a learned distribution underestimating the uncertainty - in extreme cases collapsing to a single point. Here, we propose to approximate the posterior 
%		 for intractable-likelihood models 
		 with generative networks trained by Scoring Rule minimization, an overlooked adversarial-free method enabling smooth training and better uncertainty quantification. 
%		Scoring Rules evaluate probability distributions with respect to an outcome; we employ them to assess how the approximate posterior for a simulated observation matches the corresponding parameter value.
%		Some SRs are minimized when approximate and true posteriors coincide and can be estimated via samples from the approximate posterior. This allows relaxing the invertibility requirement and using more general neural networks.
		In simulation studies, the Scoring Rule approach yields better performances with shorter training time with respect to the adversarial framework.
	\end{abstract}

	\section{Introduction}
	%\subsection{Likelihood-Free Inference}
	
	Intractable-likelihood models are models for which it is impossible to evaluate the likelihood $ p(\ddobs|\ddtheta) $ for an observation $\ddobs $, but from which it is easy to simulate for any parameter value $ \ddtheta $.
	Given $ \ddobs $ and a prior $ \pi(\ddtheta) $, the standard Bayesian posterior is $ \pi(\ddtheta|\ddobs) \propto \pi(\ddtheta) p(\ddobs|\ddtheta) $.
	However, obtaining that explicitly or sampling from it with Markov Chain Monte Carlo (MCMC) is impossible without having access to the likelihood.
	
	Likelihood-Free Inference (LFI) techniques exploit model simulations to approximate the exact posterior distribution when the likelihood is unavailable. Popular approaches include Approximate Bayesian Computation methods \citep{lintusaari2017, bernton2019approximate} and Synthetic Likelihood \citep{price2018bayesian, an2020robust}.
	% replace the intractable likelihood with a surrogate one whose parameters are estimated from simulations. The second category is constituted by , which implicitly approximate the likelihood by weighting parameter values according to the mismatch between observed and simulated data.
	
	%\subsection{Likelihood-Free Inference with Neural Networks}
	
	A recent strand of literature \citep{papamakarios2016fast, lueckmann2017flexible, papamakarios2018sequential, lueckmann2019likelihood, greenberg2019automatic, durkan2020contrastive, radev2020bayesflow} has explored using neural networks to perform LFI\footnote{Check \href{https://neurallikelihoodfreeinference.github.io/}{this website} for an interactive up-to-date list.}. Many methods employ normalizing flows \citep{papamakarios2019normalizing}: invertible neural networks which parametrize complex probability distributions by transforming a simple one (say, multivariate Gaussian). Normalizing flows allow direct sampling from the parametrized distribution and explicit density evaluation via the change-of-variables formula enabled by invertibility; using the latter, normalizing flows can be trained via maximum likelihood estimation on parameter-simulation pairs. They can be used to represent either the likelihood \citep{papamakarios2018sequential, lueckmann2019likelihood} or the posterior \citep{papamakarios2016fast, lueckmann2017flexible, greenberg2019automatic, radev2020bayesflow}.
	
	However, requiring invertibility strongly constrains the network architecture. More general generative networks drop this requirement, thus gaining expressiveness and the ability to easily scale to large input and output sizes, but forgoing density evaluation: from a generative network, you can only obtain draws from the parametrized probability distribution.
	%, instead, allow to sample from a generic probability distribution by transforming draws from a base one via a non-invertible transformation, thus foregoing exact density evaluation.
	For this reason, maximum likelihood estimation of neural network weights is impossible and people use training methods based on generating samples from the generative network. A paradigmatic example is the popular Generative Adversarial Network, or GAN, framework of \citealp{goodfellow2014generative}, where the generative network is trained in a min-max game against an additional \textit{discriminator} network aiming at distinguishing between training samples and simulations from the generative network\footnote{Notice that generative networks and adversarial training are older techniques than normalizing flows. We introduce them in this order by following their usage in the LFI context.}. 
	
	For LFI, \cite{ramesh2022gatsbi} used a generative network to represent a posterior approximation and trained it with an adversarial approach. From the trained network, approximate posterior samples can be directly obtained.
	Here, we build on \cite{ramesh2022gatsbi} by proposing a different training strategy based on minimizing values of Scoring Rules (SR, \citealp{gneiting2007strictly}), which are functions assessing the match between a probability distribution and an observation. In contrast to the adversarial approach, plagued by biased gradient estimates \citep{binkowski2018demystifying} and often leading to mode collapse (in which the parametrized probability distribution collapses on a single point, \citealp{richardson2018gans}), SR training has been found to better capture the full shape of the probability distribution in the setting of probabilistic forecasting \citep{pacchiardi2021probabilistic}.
	%	 : rather than relying on the adversarial framework,
	%, we propose to train them by  
%	Mode collapse may not be a problem in some applications of generative network not requiring careful uncertainty estimation, but it represents a large issue in case posterior approximation is required.
	Additionally, the SR-minimization approach leads to simpler training with respect to the adversarial one, as it does not require a discriminator or solving a min-max problem.
	
%	 As already discussed in \cite{pacchiardi2021probabilistic}, both the adversarial and the SR-based approaches to train generative networks can be derived from a divergence-minimization problem. 
	
	The rest of the paper is organized as follows. Section~\ref{sec:gen_net_post_inference} discusses how to use a generative network to represents and approximate posterior and reviews the training method employed in \cite{ramesh2022gatsbi}. Section~\ref{sec:SR_approach} we introduce the SR-minimization training for LFI. Section~\ref{sec:results} reports simulation results and Section~\ref{sec:conclusions} gives concluding remarks.

	\paragraph{Notation} 	
	
	We will denote respectively by $ \mathcal{Y} \subseteq \R^d $ and $ \Theta \subseteq \R^p $ the data and parameter space. %, which we assume to be Borel sets. 
	We will 
	%	assume the observations are generated by a distribution $ P_0 $ with density $ p_0(\cdot|\theta) $ and
	use $ P(\cdot|\ddtheta) $ and $ p(\cdot|\ddtheta) $ to denote the distribution and likelihood (with respect to Lebesgue measure) of the considered likelihood-free model. $ \Pi $ and $ \pi $ will denote prior distribution and prior density on $ \Theta $, and $ \Pi(\cdot|\ddobs) $ and $ \pi(\cdot|\ddobs) $ will denote corresponding posterior quantities for observation $ \ddobs $. In general, we will use $ P $ or $ Q $ to denote distributions, while $ S $ will denote a generic Scoring Rule. Other upper-case letters ($ \Ddsim,\Ddobs$ and ${\Ddz} $) will denote random variables while lower-case ones will denote observed (fixed) values. We will denote by $ \Ddobs $ or $ \ddobs $ the observations (correspondingly random variables and realizations).
	%	 and $ \Dsim $ or $ \dsim $ the simulations.
	Bold symbols denote vectors, and subscripts to bold symbols denote sample index (for instance, $ \ddobs_i $). Instead, subscripts to normal symbols denote component indices (for instance, $ y_j $ is the $ j $-th component of $ \ddobs $, and $ y_{i,j} $ is the $ j $-th component of $ \ddobs_i $).
	%Subscripts will denote sample index and superscripts vector components.
	%Also, we will respectively denote by $ \Ddobsn = \{ \Dobs_i\}_{i=1}^n \in \X^n $ and $ \ddobsn = \{ \dobs_i\}_{i=1}^n \in \X^n $ a set of random and fixed observations. Similarly, $ \Ddsimm = \{ \Dsim_j\}_{j=1}^m \in \X^m  $ and $ \ddsimm = \{ \dsim_j\}_{j=1}^m \in \X^m   $ denote a set of random and fixed model simulations. 
	Finally, $ \independent $ will denote independence between random variables, while $ \Ddobs \sim P$ indicates a random variable distributed according to $ P $ and $ \ddobs \sim P$ a sample from such random variable.% By abusing notation, sometimes we will use $ \Ddobs \sim p $ for a likelihood $ p $ in place of the corresponding distribution $ P $.
	
	%We use upper case to denote random variables, and their lower-case counterpart to denote observed values. Bold symbols denote vectors, and subscripts to bold symbols denote sample index (for instance, $ \ddobs_t $). Instead, subscripts to normal symbols denote component indices (for instance, $ y_i $ is the $ i $-th component of $ \ddobs $, and $ y_{t,j} $ is the $ j $-th component of $ \ddobs_t $). Finally, we use notation $\ddobs_{j:k} = (\ddobs_j, \ddobs_{j+1}, \ldots, \ddobs_{k-1}, \ddobs_k)$, for $j\le k$.

	\section{Approximate posterior via generative network}\label{sec:gen_net_post_inference}
	We use a generative network to represent an approximate posterior distribution $ Q_\phi(\cdot|\ddobs) $ on the parameter space $ \Theta $ given an observation $ \ddobs\in \Y $. The density of $ Q_\phi(\cdot|\ddobs) $ (with respect to the Lebesgue measure) will be denoted by $ q_\phi(\cdot|\ddobs) $. The generative network is defined via a neural network $ g_\phi : \Z \times \Y \to \Theta  $ transforming samples from a probability distribution $ P_\ddz $ over the space $ \Z $ conditionally on an observation $ \ddobs\in \Y $; $ \phi $ represents neural network weights. Samples from $ Q_\phi(\cdot|\ddobs) $ are therefore obtained by sampling $ \ddz \sim P_\ddz $ and computing  $\tilde \ddtheta  = g_\phi(\ddz,\ddobs) \sim Q_\phi(\cdot|\ddobs) $\footnote{
%		Denoting respectively by $ Q_\phi $ and $ p_\ddz $ the distributions induced by $ q_\phi $ and $ q_z $ (considering the latter two as being densities with respect to the Lebesgue measure), 
		Formally, $ Q_\phi(\cdot|\ddobs) $ is the push-forward of $ P_\ddz $ through the map $ g_\phi(\cdot, \ddobs) $: $Q_\phi(\cdot|\ddobs) = g_\phi(\cdot, \ddobs)\sharp P_\ddz $, which means that, for any set $ A $ belonging to the Borel $ \sigma $-algebra $ \sigma(\Theta) $, 
		$Q_\phi(A|\ddobs)= P_\ddz\left( \{ \ddz\in \mathcal{\Ddz} : g_\phi(\ddz, \ddobs) \in A\} \right)$.
	}

	In the following, as it is standard in the LFI setup, we assume to have access to parameter-simulations pairs $ (\ddtheta_i, \ddobs_i)_{i=1}^n $ generated from the prior $ \ddtheta_i \sim \Pi $ and the model $ \ddobs_i \sim P(\cdot|\ddtheta_i) $; critically, these can also be considered as being samples from the data marginal $ \ddobs_i \sim P $ and the posterior $ \ddtheta_i \sim \Pi(\cdot|\ddobs_i)$. Using these samples, we want to tune $ \phi $ such that $ Q_\phi(\cdot|\ddobs ) \approx \Pi(\cdot|\ddobs) $ for all values of $ \ddobs $; this is therefore an \textit{amortized} setting \citep{radev2020bayesflow}, i.e. simulations from the likelihood-free models are drawn independently from the observations on which inference is required. We discuss strategies for tailoring simulations to a specific observation in Sec~\ref{sec:sequential}.

	\begin{comment}
	A possibility is considering a statistical divergence $ D $, i.e. a function of two distributions, say $ p $ and $ q $, such that $ D(p||q)\ge 0 $ and $ D(p||q) = 0 \iff p=q  $. For a given $ D $, solving:
	\begin{equation}
	\underset{\phi}{\arg \min }\ \E_{Y \sim p(\ddobs)} D\left(\pi(\cdot|Y) \| p_\phi(\cdot|Y) \right)
	\end{equation}
	would yield $ q_\phi(\cdot|\ddobs ) = \pi(\cdot|\ddobs) $ for all values of $ \ddobs $. Solving the above problem, or even minimizing it with Stochastic Gradient Descent (SGD), does not seem to be possible in general by only using samples from $ q_\phi(\cdot|\ddobs ) $ and $ \pi(\cdot|\ddobs)$. We discuss below how the adversarial and scoring-rule based approaches develop ways to tackle this.
	%where $ p(\ddobs) $ denotes the data marginal (the evidence) under the prior $ \pi $ and likelihood $  p(\ddobs|\theta) $.
	\end{comment}

	\subsection{Adversarial posterior inference}\label{sec:adversarial}
	
	\begin{comment}
	We consider now $ D $ to be an \textit{f}-divergence: 
	\begin{equation}
	D_{f}\left(p \| q\right)=\int_{\mathcal{Y}} q(\ddobs) f\left(\frac{p(\ddobs)}{q(\ddobs)}\right) d \ddobs,
	\end{equation}
	where $ f:\R_+ \to\R $  is a convex, lower-semicontinuous function for which $ f(1)=0 $. 
	\end{comment}

	In \cite{ramesh2022gatsbi}, the posterior approximation $ Q_\phi $ is trained in an adversarial framework. 
%	as we can only generate draws from $ q_\phi  $, 
	This requires introducing a \textit{discriminator} or \textit{critic} neural network $ c_\psi:\Theta\times \Y\to \R $ with weights $ \psi $  whose task is to distinguish draws from the approximate and true posteriors. The loss employed in \cite{ramesh2022gatsbi} is the conditional version of the original GAN loss from \cite{goodfellow2014generative}, which was originally discussed in \cite{mirza2014conditional}: 
	\begin{equation}\label{Eq:gan_loss}
	\begin{aligned}
	L(\phi, \psi) &=\mathbb{E}_{\ddtheta \sim \Pi}\E_{\Ddobs\sim P(\cdot \mid \ddtheta)}\E_{\Ddz \sim P_\ddz}\left[\log c_{\psi}(\ddtheta, \Ddobs)+\log \left(1-c_{\psi}\left(g_{\phi}(\Ddz, \Ddobs), \Ddobs\right)\right)\right] \\
	&=\mathbb{E}_{\Ddobs \sim P}\left[\mathbb{E}_{\ddtheta \sim \Pi(\cdot\mid \Ddobs )}\left(\log c_{\psi}(\ddtheta, \Ddobs )\right)+\mathbb{E}_{\tilde \ddtheta \sim Q_\phi(\cdot\mid \Ddobs)}\left(\log \left(1-c_{\psi}(\tilde \ddtheta, \Ddobs)\right)\right)\right],
	\end{aligned}
	\end{equation}
	whose saddle point solution
	\begin{equation}\label{Eq:gan}
	\min_{\phi} \max_{\psi} L(\phi, \psi)
	\end{equation}
	leads to $ Q_\phi(\cdot|\ddobs) $ being the exact posterior for all choices of $ \ddobs $ for which $ p(\ddobs)>0 $ (provided $ q_\phi $ and $ c_\psi $ have infinite expressive power; that in fact corresponds to minimizing the Jensen-Shannon divergence, see Appendix~\ref{app:f-GAN} for more details). 
	
	In practice, the training procedure works by Stochastic Gradient Descent (SGD): we replace the expectations in Eq.~\eqref{Eq:gan_loss} with empirical means over (a mini-batch of) the training dataset and draws from the generative network and alternate maximization steps over $ \psi $ with minimization steps over $ \phi $. This alternating optimization is however unstable and requires careful hyperparameters tuning and specialized training routines; additionally, doing a finite number of maximization steps over $ \psi $ and then using the current value of $ \psi $ to compute gradients of the objective with respect to $ \phi $ leads to biased gradient estimates \cite{binkowski2018demystifying}. A possible consequence of this is mode collapse \cite{richardson2018gans}, in which the distribution parametrized by the generative network collapses onto a single point. This may not be an issue in some applications of generative networks where uncertainty quantification is not important, but it can be detrimental for approximate posterior inference.

	\section{Posterior inference via Scoring Rules minimization}\label{sec:SR_approach}
	
	We discuss here how to use Scoring Rules to define an adversarial-free training objective for generative networks, focusing on the specific case of a generative network parametrizing an approximate posterior. In \cite{pacchiardi2021probabilistic}, more details on SR-training and its application to probabilistic forecasting can be found. Other works employing SR training, albeit not for the LFI framework, are \cite{bouchacourt2016disco, gritsenko2020spectral, harakeh2021estimating}.
	
	\subsection{Scoring Rule training}
	We first introduce Scoring Rules for a distribution $ P $ related to a generic random variable $ \Ddsim $. A Scoring Rule (SR, \citealp{gneiting2007strictly}) $ S(P, \ddsim) $ is a function of $ P $ and of an observation $ \ddsim $ of the random variable $ \Ddsim $. If $ \Ddsim $ is actually distributed according to $ Q $,
%	If $\ddsim$ is a realization of a random variable , 
	the expected Scoring Rule is defined as:
	\begin{equation*}
	S(P,Q) := \E_{\Ddsim \sim Q} S(P, \Ddsim),
	\end{equation*}
	The Scoring Rule $ S $ is \textit{proper} relative to a set of distributions $ \mathcal{P} $ over $ \X $ if $$ S(Q, Q) \le S(P,Q) \ \forall \ P,Q \in \mathcal{P},$$ i.e., if the expected Scoring Rule is minimized in $ P $ when $ P=Q $. Moreover, $ S $ is \textit{strictly proper} relative to $ \mathcal{P} $ if $ P = Q $ is the unique minimum: $$ S(Q,Q) < S(P,Q)  \ \forall \ P, Q \in \mathcal{P}  \text{ s.t. }  P\neq Q.$$
	
	Let us now go back to the Bayesian LFI setting introduced at the start of the paper.
	Denoting by $ Q_\phi(\cdot|\ddobs) $ the approximate posterior parametrized by the generative network, solving the following problem for a strictly proper $ S $:
	\begin{equation}\label{Eq:SR_obj_cond}
	\argmin_\phi \E_{\Ddobs\sim P} \E_{\ddtheta\sim\Pi(\cdot|\Ddobs)}  S(Q_\phi(\cdot|\Ddobs),\ddtheta) = 	\argmin_\phi \E_{\ddtheta\sim\Pi} \E_{\Ddobs\sim P(\cdot|\ddtheta)}  S(Q_\phi(\cdot|\Ddobs),\ddtheta)
	\end{equation}
	leads to $q_\phi(\cdot|\ddobs) =\pi(\cdot|\ddobs) $ for all values of $ \ddobs $ for which $ p(\ddobs)>0 $.
	
	%, Eq.~\eqref{Eq:SR_obj_cond} is minimized when . 
	
	Replacing the expectations in Eq.~\eqref{Eq:SR_obj_cond} with empirical means over the training dataset yields: 
	%The objective in Eq.~\eqref{Eq:SR_obj_cond_emp} is an empirical estimate of:
	\begin{equation}\label{Eq:SR_obj_cond_emp}
	\argmin_\phi \frac{1}{n} \sum_{i=1}^n S(Q_\phi(\cdot|\ddobs_i),\ddtheta_i);
	\end{equation}
	computing the objective directly is intractable as, in general, we do not have access to $ S(Q_\phi(\cdot|\ddobs),\ddtheta) $. Notice, however, that in order to train $ Q_\phi$ via SGD it is enough to obtain unbiased estimates of $ \nabla_\phi S(Q_\phi(\cdot|\ddobs_i),\ddtheta_i) $, which can be easily done whenever $ S $ admits an easy unbiased empirical estimator $ \hat S $, i.e. such that: 
	\begin{equation}\label{}
	\E \left[\hat S(\{\tilde \ddtheta_j^{(\ddobs)}\}_{j=1}^m, \ddtheta)\right] = S(Q_\phi(\cdot|\ddobs),\ddtheta),
	\end{equation}
	where the expectation is over $ \tilde \ddtheta_j^{(\ddobs)} \sim Q_\phi(\cdot|\ddobs) $. More details can be found in Appendix~\ref{app:unbiased_estimate_obj}.
	%, then gradient with respect to $ \phi $ of:
	%\begin{equation}\label{}
	%\frac{1}{n} \sum_{i=1}^n S(\{\tilde \theta_j^{(\ddobs)}\}_{j=1}^m, \theta);
	%\end{equation}
	If $ S $ admits such an estimator, each step of SGD involves generating $ m $ simulations from the generative network $ Q_\phi(\cdot|\ddobs_i) $ for each $ \ddobs_i $ in the training batch.
%	Therefore, training a generative network via SR minimization requires choosing a SR $ S $ for which unbiased gradient estimates are easily available and running SGD on the objective in
%	 Eq~\eqref{Eq:SR_obj_cond_emp}, where at each step . 

	Below, we introduce some SRs for which easy unbiased estimators of the form above are available. These estimators however require $ m >1$; to train GAN, instead, a single draw from the generative network was enough. In experiments, however, small values of $ m $ ($ m=10 $ for instance) lead to satisfactory results. Additionally, as mentioned above, the SR approach does not require a discriminator network and has a smoother training process, which implies convergence is generally reached with less training epochs. These two factors lead to lower computational and memory cost with respect to adversarial training (see Section~\ref{sec:results} for details).
	
%	Finally, from a practical standpoint, training two neural networks has a larger memory and computational footprint.

	\subsection{Some Scoring Rules with unbiased estimators}

	We introduce two specific Scoring Rules that we use in our experiments, by considering again a generic distribution $ P $ and random variable $ \Ddsim $.

	\paragraph{Energy score}
	The energy score\footnote{The probabilistic forecasting literature \citep{gneiting2007strictly} use a different convention for the energy score and the subsequent kernel score, which amounts to multiplying our definitions by $ 1/2 $. We follow here the convention used in the statistical inference literature \citep{rizzo2016energy, cherief2020mmd, nguyen2020approximate}}.  is given by:
	\begin{equation}\label{Eq:eng_score}
	\SE^{(\beta)}(P, \ddsim) = 2 \cdot \E \left[\| \tilde  \Ddsim - \ddsim\|_2^\beta\right] - \E\left[\|\tilde  \Ddsim- \tilde  \Ddsim'\|_2^\beta\right] ,\quad  \tilde  \Ddsim \independent  \tilde  \Ddsim' \sim P,
	\end{equation}
	where $ \beta \in (0,2) $.
	This is a strictly proper SR for the class 
%	$ \mathcal{P}_\beta (\X)$
	 of probability measures $ P $ such that $ \E_{\tilde  \Ddsim\sim P}\|\tilde  \Ddsim\|^\beta < \infty $ \citep{gneiting2007strictly}.
	%The related divergence is the square of the energy distance, which is a metric between probability distributions (\citealt{rizzo2016energy}; see Appendix~\ref{app:energy})
	An unbiased estimate can be obtained by replacing the expectations in $\SE^{(\beta)}$ with empirical means over draws from $ P $ (see Appendix~\ref{app:unbiased_SR})
	%unbiasedly estimating the expectations in $\SE^{(\beta)}$ :
	%\begin{equation}
	%\hat S_{\text{E}}^{(\beta)}(\tilde  \ddsimm, \dobs) =\frac{2}{m} \sum_{j=1}^m \left\| \dsim_j - \dobs\right\|_2^\beta - \frac{1}{m(m-1)}\sumjk \left\|\dsim_j-\dsim_k\right\|_2^\beta.
	%\end{equation}
	We will fix $ \beta=1$ in the rest of this work. % and we will write $ \SE $ in place of $ \SE^{(1)} $.

	\paragraph{Kernel score}
	When $ k(\cdot, \cdot) $ is a positive definite kernel, the kernel score for $ k $ can be defined as \citep{gneiting2007strictly}:
	\begin{equation}\label{Eq:kernel_score}
	S_k(P, \ddsim) = \E[k(\tilde  \Ddsim,\tilde  \Ddsim')] - 2\cdot\E [k(\tilde  \Ddsim, \ddsim)],\quad  \tilde  \Ddsim \independent  \tilde  \Ddsim' \sim P.
	\end{equation}
	%The corresponding divergence is the squared Maximum Mean Discrepancy (MMD, \citealp{gretton2012kernel}) relative to $ k $ (see Appendix~\ref{app:MMD}).
	The kernel score is proper for the class of probability distributions  $ P $ for which $ \E_{\tilde  \Ddsim,\tilde  \Ddsim' \sim P}[k(\tilde  \Ddsim,\tilde  \Ddsim')] $ is finite (by Theorem 4 in \cite{gneiting2007strictly}). It is closely related to the kernel Maximum Mean Discrepancy (MMD, \citealp{gretton2012kernel}) and is strictly proper under conditions ensuring the MMD is a metric \citep{gretton2012kernel}. These conditions are satisfied, among others, by the Gaussian kernel (which we will use in this work):
	\begin{equation}\label{Eq:gau_k}
	k(\tilde  \ddsim, \ddsim)=\exp \left(-\frac{\|\tilde  \ddsim-\ddsim\|_{2}^{2}}{2 \gamma^{2}}\right),
	\end{equation}
	in which $ \gamma $ is a scalar bandwidth.
	As for the Energy Score, an unbiased estimate can be obtained by replacing the expectations in $S_k$ with empirical means over draws from $ P $ (see Appendix~\ref{app:unbiased_SR}).
	
	%Similarly to the Energy Score, an unbiased
	%estimate is:
	%\begin{equation}
	%\hat S_k(\ddsimm, \dobs) = \frac{1}{m(m-1)}\sumjk  k(\dsim_j,\dsim_k )-\frac{2}{m} \sum_{j=1}^m k(\dsim_j,\dobs).
	%\end{equation}

	\paragraph{Patched SR}
	
	We now discuss a way to build a composite SR which encodes structural information in $ \Ddsim $. 	
	In fact, if $ \Ddsim $ has some structure (say, it is on a 1D or 2D grid), computing the raw SRs above discards that information. A way to encode some of it is to compute the SRs on localized \textit{patches} across the grid and cumulate the score; in this way, short-scale correlations are given more importance. However, the resulting Scoring Rule is not strictly proper; to fix this, we add the SR computed over the full $ \ddsim $, which makes the overall SR strictly proper (see Lemma 3.4 in \cite{pacchiardi2021probabilistic}). 
	
	For a given SR $ S $, therefore, the patched SR is: 
	\begin{equation}\label{Eq:patched_SR}
	S_{p}(P, \ddsim) = w_1 S(P, \ddsim) + w_2 \sum_{p \in \mathcal{P}} S(P|_p, \ddsim|_p),
	\end{equation}
%	\begin{equation}\label{Eq:patched_SR}
%S_{p}(Q_\phi(\cdot|\ddobs), \ddsim) = w_1 S(Q_\phi(\cdot|\ddobs), \ddtheta) + w_2 \sum_{p \in \mathcal{P}} S(Q_\phi(\cdot|\ddobs)|_p, \ddtheta|_p),
%\end{equation}
	where $ w_1, w_2 >0 $, $ |_p$ denotes the restriction of a distribution or of a vector to a patch $p $ and $ \mathcal{P} $ is a set of patches. See \cite{pacchiardi2021probabilistic} for more discussion on patched SRs.

	\subsection{Connection with normalizing flows}
	
	As mentioned in the introduction, normalizing flows are generative networks which impose invertibility of the map $ g_\phi(\ddz,\ddobs) $ with respect to $\ddz $. As such, density evaluation of the resulting $ q_\phi $ is possible via the change-of-variables formula, so that $ \phi $ is usually trained via maximum likelihood \citep{papamakarios2019normalizing}. For instance, in \cite{radev2020bayesflow}, the following problem is considered, where $ \mathbb{K} \mathbb{L} $ denotes the Kullback-Leibler divergence: 
	\begin{equation}
	\begin{aligned}
%	\widehat{\boldsymbol{\phi}}
	&\quad \ \underset{\boldsymbol{\phi}}{\operatorname{argmin}}\ \mathbb{E}_{\Ddobs \sim P}\left[\mathbb{K} \mathbb{L}\left(\Pi(\cdot \mid \Ddobs) \| Q_{\boldsymbol{\phi}}(\cdot \mid \Ddobs)\right)\right]\\
	&=\underset{\phi}{\operatorname{argmin}} \mathbb{E}_{\Ddobs \sim P} \mathbb{E}_{\boldsymbol{\ddtheta} \sim \Pi(\cdot|\Ddobs)} \left[ - \log q_{\boldsymbol{\phi}}(\boldsymbol{\ddtheta} \mid \Ddobs) \right] \\
	&=\underset{\phi}{\operatorname{argmin}} \mathbb{E}_{\boldsymbol{\ddtheta} \sim \Pi} \mathbb{E}_{\Ddobs \sim P(\cdot|\boldsymbol{\ddtheta})} \left[ - \log q_{\boldsymbol{\phi}}(\boldsymbol{\ddtheta} \mid \Ddobs) \right],
	\end{aligned}
	\end{equation}
	which corresponds to our SR-based approach in Eq.~\eqref{Eq:SR_obj_cond} by identifying $ S(Q_\phi(\cdot|\ddobs),\ddtheta) = -\log q_\phi(\ddtheta|\ddobs) $, which is the strictly-proper logarithmic scoring rule. %he normalizing flows approaches can be cast as a SR-minimization approach. 

	\subsection{Sequential training}\label{sec:sequential}
	
	Up to this point, we have considered the training data from the simulator model $ (\ddtheta_i, \ddobs_i)_{i=1}^n  $ to be generated independently from the observation on which inference is performed; under this assumption, we have discussed ways to learn posterior approximations valid for all values of $ \ddobs $ such that $ p(\ddobs)>0 $. Once the neural network is trained, therefore, inference can be performed for as many observations as we wish. This is a so-called \textit{amortized} setup \cite{radev2020bayesflow}.
	
	However, practitioners may require posterior inference for a single observation $ \ddobs_o $. What they are interested in, therefore, is the quality of the approximation for values of $ \ddtheta $ at which the true posterior density is large for the observed $ \ddobs_o $.
%	close to the observed $ \ddobs $. 
	In this case, generating training samples independently from $ \ddobs_o $ may be wasteful: a more efficient method (in terms of simulations from the model $ p(\cdot|\ddtheta) $) would generate training samples $ \ddtheta_i $'s close to the modes of the true posterior.
%		a better use of computin as those far away from the observed value contain little information about the shape of the posterior close to it. 
%	Therefore, it may be more efficient  to tailor simulation close to the observed $ \ddobs $. 
	This can be done in a sequential fashion: given a small amount of training data, a first approximation $ Q_{\phi_1} $ is obtained; from that, additional training samples $ (\ddtheta_i, \ddobs_i) $ are generated by $ \ddtheta_i \sim Q_{\phi_1}(\cdot|\ddobs_o), \ddobs_i \sim P(\cdot|\ddtheta_i)$ and used to (re-)train an approximation $ Q_{\phi_2} $. This procedure is iterated several times, allowing therefore the training samples to progressively better cover the posterior modes, which in turn allows to refine the posterior approximation close to the modes \citep{lueckmann2017flexible, greenberg2019automatic}.
	
	However, naively following that strategy is incorrect. To see this, assume that, at the second round, we just train on samples drawn from the approximate posterior $ \tilde \Pi= Q_{\phi_1}(\cdot|\ddobs_o) $ obtained at the first round. Such a sampled pair $ (\ddtheta_i, \ddobs_i) $ was drawn from a joint density
%	$\tilde p(\ddtheta_i, \ddobs_i) = 
	$\tilde \pi(\ddtheta_i)p(\ddobs_i|\ddtheta_i) = \tilde p(\ddobs_i) \tilde \pi(\ddtheta_i|\ddobs_i),$ where $ \tilde \pi $ on the left-hand side of the equality is the density of the proposal $ \tilde \Pi $ and the quantities on the right-hand side are univocally defined by the left-hand side. The optimal $ \phi^{\star} $ obtained via SR-minimization thus corresponds to 
	%Clearly, you need to fix the training objective in some way as otherwise you would get 
	$q_{\phi^{\star}}(\cdot|\ddobs)= \tilde \pi(\cdot|\ddobs)$, which is not the correct target. 
	
	%if you just replace $\pi$ with $\tilde \pi$ in the original objective.

	The traditional way to fix this entails introducing importance weights in the training objective (Eq.~\ref{Eq:SR_obj_cond}):
	\begin{equation}\label{key}
	%L(\phi) =
%	 \E_{\Ddobs\sim P} \E_{\ddtheta\sim\Pi(\cdot|\Ddobs)}  S(Q_\phi(\cdot|\Ddobs),\ddtheta) = 
	 \E_{\ddtheta\sim \Pi} \E_{\Ddobs\sim P(\cdot|\ddtheta)}  S(Q_\phi(\cdot|\Ddobs), \ddtheta) = \E_{\ddtheta\sim \tilde \Pi} \frac{\pi(\ddtheta)}{\tilde\pi(\ddtheta)} \E_{\Ddobs\sim P(\cdot|\ddtheta)}   S(Q_\phi(\cdot|\Ddobs), \ddtheta).
	\end{equation}
	As $ \tilde\pi(\ddtheta) $ cannot be evaluated, a solution is to fit a probabilistic classifier (at each round of the sequential procedure) to samples from $ \pi(\ddtheta) $ and $ \tilde\pi(\ddtheta) $ and use it to estimate the ratio $ \frac{\pi(\ddtheta)}{\tilde\pi(\ddtheta)} $. This classifier is not required for the normalizing flows approaches, where the ratio can be evaluated explicitly \citep{lueckmann2017flexible, greenberg2019automatic} (unless the prior $ \pi $ is also defined implicitly, as in the camera model example in Section~\ref{sec:results}). For the GAN approach, a similar importance weights approach requires additionally to estimate the ratio $ \frac{\tilde p(\ddobs)}{p(\ddobs)} $ \citep{ramesh2022gatsbi}.
	
	%Notice how, in contrast to the importance weights in GATSBI, here you have a single ratio to be estimated. 

	An alternative approach, which was proposed in \cite{ramesh2022gatsbi}, involves correcting the distribution of the variable $ \Ddz $ which is transformed by the generative network. Specifically, \cite{ramesh2022gatsbi} shows that $\pi(\ddtheta|\ddobs) = \tilde \pi(\ddtheta|\ddobs) w(\ddtheta,\ddobs) \iff \tilde \pi(\ddtheta|\ddobs) =  \pi(\ddtheta|\ddobs) (w(\ddtheta,\ddobs))^{-1}$, where $w(\ddtheta,\ddobs) = \frac{\pi(\ddtheta)}{\tilde\pi(\ddtheta)} \frac{\tilde p(\ddobs)}{p(\ddobs)}$. Therefore you can consider a modified approximation $\tilde Q_\phi(\cdot|\Ddobs)$ and a new training objective:
	\begin{equation}\label{Eq:sequential}
	\E_{\Ddobs\sim \tilde P} \E_{\ddtheta\sim \tilde \Pi(\cdot|\Ddobs)} S(\tilde Q_\phi(\cdot|\Ddobs), \ddtheta) = \E_{\ddtheta\sim \tilde \Pi} \E_{\Ddobs\sim P(\cdot|\ddtheta)}  S(\tilde Q_\phi(\cdot|\Ddobs), \ddtheta)
	\end{equation}
	whose minimization leads to $\tilde Q_\phi(\cdot|\Ddobs) = \tilde \Pi(\cdot|\Ddobs)$. By setting $$\tilde Q_\phi(\cdot|\Ddobs) = Q_\phi(\cdot|\Ddobs) (w(\ddtheta,\ddobs))^{-1},$$ you ensure  $Q_\phi(\cdot|\Ddobs) = \Pi(\cdot|\Ddobs)$. 
	To train $ \phi $ using the objective in Eq.~\eqref{Eq:sequential}, draws from $\tilde Q_\phi(\cdot|\Ddobs)$ are required; those can be obtained by sampling $\ddz\sim \tilde P_\ddz$ where $\tilde p_\ddz(\ddz) = p_\ddz(\ddz) (w(g_\phi(\ddz,\ddobs),\ddobs))^{-1}$ and then computing $\ddtheta = g_\phi(\ddz,\ddobs)$, which is thus a sample from $\tilde Q_\phi(\cdot|\Ddobs)$. Sampling from $ \tilde P_\ddz $ entails either Rejection sampling or MCMC. Additionally, this method, this correction requires estimating two ratios via probabilistic classifiers ($  \frac{\tilde p(\ddobs)}{p(\ddobs)}  $ and $ \frac{\pi(\ddtheta)}{\tilde\pi(\ddtheta)}  $). The advantage with respect to the importance weight strategy is reduced variance of the training objective; however, a larger computational cost in obtaining the corrected samples from the latent distribution is involved (with repeated passes through the NNs approximating the ratios for each training sample $ \ddobs_i $). The strategy discussed above can seamlessly be applied in the GAN approach as well \citep{ramesh2022gatsbi}.

	On the considered examples in \cite{ramesh2022gatsbi}, the sequential approaches did not provide any advantage with respect to the amortized ones, mainly due to the additional computational cost associated to estimating the ratios. As we use the same examples here, we do not test these methods, but we discussed them anyway as they may turn out to be useful in other applications.

	%1. Need to train one or two classifiers NNs to estimate the marginal ratios, at each training round
	%2. Need to run MCMC or rejection sampling to sample from the latent $Z$ in the second approach; that involves repeated passes through the NN, for each simulation output. Additionally that scales linearly with the number of simulations used to estimate the SR gradient (as in fact need to do that for each simulator output), so it is very bad. 

	\section{Simulation study}\label{sec:results}
	
	Following \cite{ramesh2022gatsbi}, we present here results on two benchmark problems and two high-dimensional models, one of which has an implicitly defined prior. For all examples, we evaluate the performance of the different methods as in \cite{ramesh2022gatsbi}. Besides that, we assess the calibration of the approximate posteriors by the discrepancy between credible intervals in the approximate posteriors and the frequency with which the true parameter belongs to the credible interval itself (we term this metric \textit{calibration error}). We also evaluate how close the posterior means are to the true parameter value by the \textit{Normalized Root Mean-Square Error} (NRMSE) and the \textit{coefficient of determination} R$^2  $; these metrics were used for LFI in \cite{radev2020bayesflow}; we provide more detail in Appendix~\ref{app:metric}. As all these metrics are for scalar variables, we compute their values independently for each component of $ \ddtheta $ and report their average.
	
	We compare our generative networks trained with SRs with the GAN-based one in \cite{ramesh2022gatsbi}; in both setups, we adapt the generative networks defined in \cite{ramesh2022gatsbi} for the different tasks. Additional training details for all models are reported in Appendix~\ref{app:training_details}. Notice how \cite{ramesh2022gatsbi} compared with additional LFI methods, concluding that the generative-network based one performs worse for the simple models but is competitive for the high-dimensional ones. %\footnote{Additionally this is the only applicable method when the prior is implicitly defined, as in the camera model example.}.
	 Here, we do not compare with these other methods as the focus of our paper is to provide a different training strategy for the generative-network based one, for which the adversarial strategy was previously the only option.

	\subsection{Benchmark models}
	
	We consider here the ``Simple Likelihood Complex Posterior'' (SLCP) and the ``Two Moons'' benchmarks; in the former, a 5-dimensional $ \ddtheta $ defines the distribution of an 8-dimensional Gaussian $ \ddobs $ in a nonlinear manner. In the Two Moons model, both $ \ddobs  $ and $ \ddtheta $ are 2-dimensional. We refer to \cite{ramesh2022gatsbi} and references therein for more details\footnote{These models are implemented in the \texttt{sbibm} Python package, whose accompanying paper \cite{lueckmann2021benchmarking} provides additional details.}
	For both models, we train all methods on $ n_\text{train} =1000, 10000 \text{ and } 100000$ posterior samples. We consider the SR methods with the Energy and Kernel Score trained with $ m= 3, 5, 10 \text{ or } 20$ samples from the generative network for each $ \ddobs_i $ in a training batch. The SR methods are trained on a single CPU, while GAN is trained on an NVIDIA Tesla-V100 GPU. For the Two Moons model, we do not use early stopping for the SR methods; additionally, we employ the optimal configuration found in \cite{ramesh2022gatsbi} for GAN. 
	
	For these two models, samples from reference posteriors are available \citep{lueckmann2021benchmarking}; therefore, as done in \cite{ramesh2022gatsbi}, we assess the performance of the different methods via the discrimination ability of a classifier trained to distinguish samples from the reference and approximate posteriors (classification-based two-sample test, C2ST). If the classification accuracy is 0.5, the classifier is unable to distinguish between the two sets of samples, implying perfect posterior approximation.
	
\begin{figure}
		\centering
		\includegraphics[width=\linewidth]{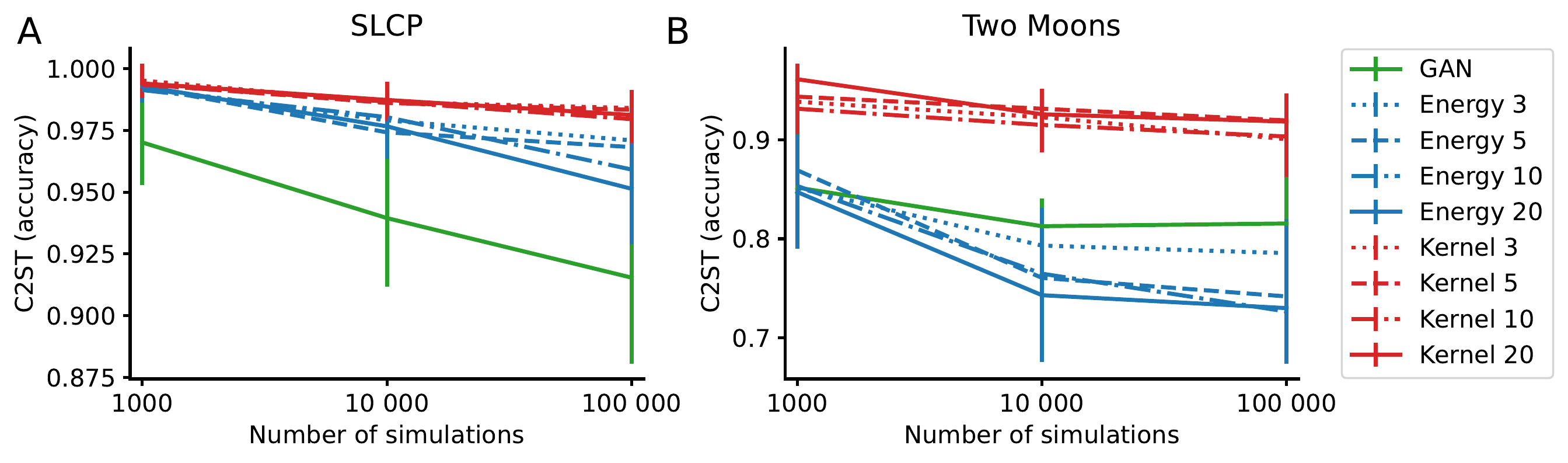}
		\caption{C2ST for SR and GAN methods for the SLCP and Two Moons benchmarks. For the SR methods, we report values for all numbers of generative network samples $ m $ used in training. SLCP: GAN performs better, but poorly on an absolute scale. Two Moons: methods based on the Energy Score perform better.}
		\label{fig:c2stfigure}
	\end{figure}

	In Figure~\ref{fig:c2stfigure}, we report C2ST values for the GAN and SR methods for the different number of training simulations. For SLCP, GAN performs better (although the performance is poor on an absolute scale and worse than other LFI methods, see \citealp{ramesh2022gatsbi}). For the Two Moons method, methods based on the Energy Score perform better.%, especially for larger number of training simulations. 
	
	In Tables~\ref{tab:SLCP_main} and \ref{tab:TwoMoons_main}, we report other performance metrics, together with the runtime and the epoch at which training was early stopped, for GAN, Energy and Kernel Score, with $ n_\text{train} =100000$ and $ m=20 $. Notice how the SR methods were trained in much shorter time (and on a single CPU). Additional results are reported in Appendices~\ref{app:res_slcp} and \ref{app:res_two_moons}.
	
	% C2ST for benchmark models

	\begin{table}[htb]
		\caption{SLCP: performance metrics, runtime and early stopping epoch for GAN, Energy and Kernel Score methods, with $ n_\text{train} =100000$ and $ m=20 $. Notice how the SR methods were trained on a single CPU, while GAN was trained on a GPU. The maximum number of training epochs was 20000.}% Cal. error and NRMSE: the smaller, the better. R$ ^2 $: the larger, the better.}
		\label{tab:SLCP_main}
		\begin{center}
			\begin{adjustbox}{max width=\textwidth}
			\begin{tabular}{lccccrr}
\hline
        & C2ST $ \downarrow $   & NRMSE $ \downarrow $   & Cal. Err. $ \downarrow $   & R$^2$ $ \uparrow $   &   Runtime (sec) &   Early stopping epoch \\
\hline
 GAN    & 0.92 $\pm$ 0.03       & 0.23 $\pm$ 0.05       & 0.06 $\pm$ 0.03            & 0.35 $\pm$ 0.30      &           30963 &                  20000 \\
 Energy & 0.95 $\pm$ 0.02       & 0.22 $\pm$ 0.06       & 0.07 $\pm$ 0.04            & 0.38 $\pm$ 0.32      &            1645 &                   2100 \\
 Kernel & 0.98 $\pm$ 0.01       & 0.22 $\pm$ 0.06       & 0.13 $\pm$ 0.10            & 0.37 $\pm$ 0.31      &            1210 &                   1200 \\
\hline
\end{tabular}
			\end{adjustbox}
		\end{center}
	\end{table}

	\begin{table}[htb]
		\caption{Two Moons: performance metrics, runtime and early stopping epoch for GAN, Energy and Kernel Score methods, with $ n_\text{train} =100000$ and $ m=20 $. Notice how the SR methods were trained on a single CPU, while GAN was trained on a GPU. Here, no early stopping was used (the maximum number of training epochs was 20000).}% Cal. error and NRMSE: the 	
		\label{tab:TwoMoons_main}
		\begin{center}
			\begin{adjustbox}{max width=\textwidth}
				\begin{tabular}{lccccrr}
\hline
        & C2ST $ \downarrow $   & NRMSE $ \downarrow $   & Cal. Err. $ \downarrow $   & R$^2$ $ \uparrow $   &   Runtime (sec) &   Early stopping epoch \\
\hline
 GAN    & 0.82 $\pm$ 0.07       & 0.20 $\pm$ 0.00       & 0.07 $\pm$ 0.02            & 0.51 $\pm$ 0.01      &           30232 &                  20000 \\
 Energy & 0.73 $\pm$ 0.04       & 0.20 $\pm$ 0.00       & 0.03 $\pm$ 0.00            & 0.51 $\pm$ 0.01      &           10805 &                  20000 \\
 Kernel & 0.92 $\pm$ 0.02       & 0.20 $\pm$ 0.00       & 0.03 $\pm$ 0.01            & 0.50 $\pm$ 0.01      &           10902 &                  20000 \\
\hline
\end{tabular}
			\end{adjustbox}
			\end{center}
	\end{table}

	\subsection{Shallow water model}
	
	The shallow water model is obtained as the discretization of a PDE describing the propagation of an initial disturbance across the surface of a shallow basin; the parameter $ \ddtheta \in \mathbb{R}^{100}$ represents the depth of the basin at equidistant points; the simulator outputs the evolution over 100 time-steps (producing a raw observation of size $ 100 \times 100 =10000$); then, a Fourier transform is computed and the real and imaginary parts are concatenated and summed to Gaussian noise, leading to $ \ddobs\in\mathbb{R}^{20k} $. More details are given in \cite{ramesh2022gatsbi}. Besides the GAN method, we test here the Energy and Kernel score with $ m=10 $ computed in three different configurations: 1) on the full parameter space, 2) with patch size 10 and step 5, and 3) with patch size 20 and step 10. Training is done on 100k samples on a NVIDIA Tesla-V100 GPU; additional details are discussed in Appendix~\ref{app:details_shallow_water}. Among the SR methods, the Energy Score with patch size 20 and step 10 performed better; therefore, we report only results for that method in the main body of the paper; results for the other configurations are given in Appendix~\ref{app:res_shallow_water}.
	
	In Figure~\ref{fig:shallow_water_model_best}, we report posterior and posterior predictive samples for both methods, together with prior samples and the ground-truth depth profile. For the Energy Score, posterior samples better follow the ground truth profile and, similarly, posterior predictive samples better match the true observation.
	
	% results for shallow water model
	\begin{figure}
		\centering
		\includegraphics[width=\linewidth]{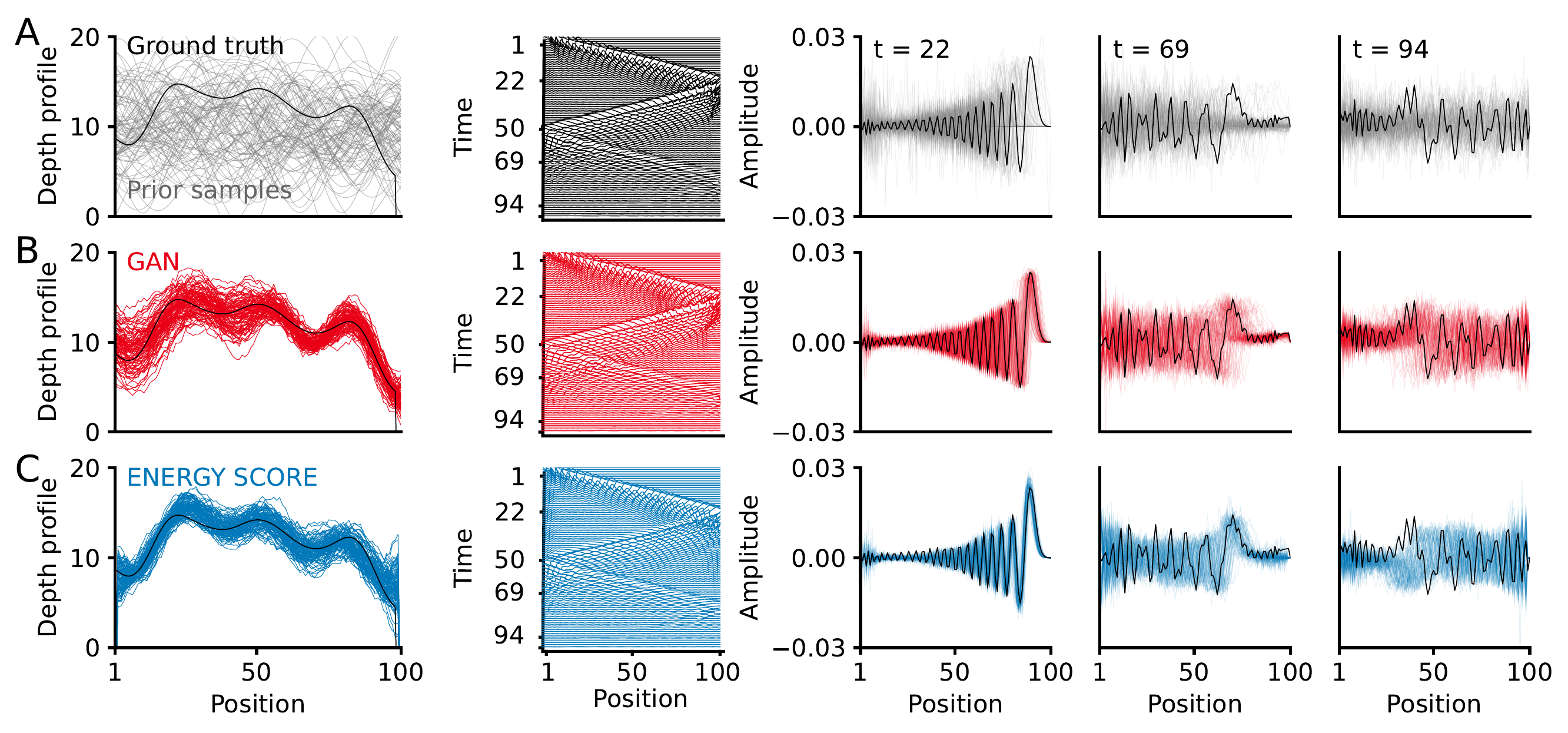}
		\caption{Shallow water model: inference results with GAN and Energy Score with patch size 20 and step 10. The figure structure closely follows that in \cite{ramesh2022gatsbi}. Row A: Ground truth, observation and prior samples. Left: ground-truth depth profile and prior samples. Middle: surface wave simulated from ground-truth profile as a function of position and time. Right: wave amplitudes at three different fixed times for ground-truth depth profile (black), and waves simulated from multiple prior samples (gray). Rows B and C refer respectively to GAN and Energy Score (with patch size 20 and step 10). For both, left represents posterior samples versus ground-truth depth profile (black), from which it can be seen how posterior samples for the Energy Score better follow the truth with respect to GAN; middle represents surface wave simulated from a single posterior sample; right represents wave amplitudes simulated from multiple posterior samples, at three different fixed times, with black line denoting the actual observation; again, Energy Score better follows the observation, except for $ t=94 $.}
		\label{fig:shallow_water_model_best}
	\end{figure}
	
	In Table~\ref{tab:shallow_water_model}, we report the performance metrics, runtime and epoch of early stopping of the GAN and Energy Score method; notice how the calibration error is much smaller for the latter, which was additionally trained in much shorter time. We also assess calibration via Simulation Based Calibration (\citealp{talts2018validating}, details in Appendix~\ref{app:sbc}) in Figure~\ref{fig:SBC_shallow_water}. That as well highlights how the calibration of the Energy Score method is better than the one achieved by GAN.
	
	\begin{table}[htb]
		\caption{Shallow Water model: performance metrics, runtime and early stopping epoch for GAN and the Energy Score with patch size 20 and step 10. The latter method achieved better results with shorter training time. We do not train GAN from scratch but rather relied on the trained network obtained in \cite{ramesh2022gatsbi}. The training time we report here corresponds to what is mentioned in \cite{ramesh2022gatsbi}, which used two GPUs for training (with respect to a single one for the SR methods). For the same reason, we do not report the epoch at which GAN training was early stopped.}% Cal. error and NRMSE: the 	
		\label{tab:shallow_water_model}
		\begin{center}
			\begin{tabular}{lcccrr}
\hline
                          & NRMSE $ \downarrow $   & Cal. Err. $ \downarrow $   & R$^2$ $ \uparrow $   &   Runtime (sec) &   Early stopping epoch \\
\hline
 Energy & 0.05 $\pm$ 0.01       & 0.03 $\pm$ 0.02            & 0.89 $\pm$ 0.05      &           60017 &                  12400 \\
GAN & 0.07 $\pm$ 0.01   & 0.12 $\pm$ 0.09   & 0.78 $\pm$ 0.05 & $ \approx $345600 & - \\
\hline
\end{tabular}
		\end{center}
	\end{table}

	% SBC lines for shallow water model; would need to compare with the GAN one as well. 
	\begin{figure}[ht]
		\centering
		\begin{subfigure}[t]{0.5\textwidth}
			\begin{center}
				\includegraphics[width=\columnwidth]{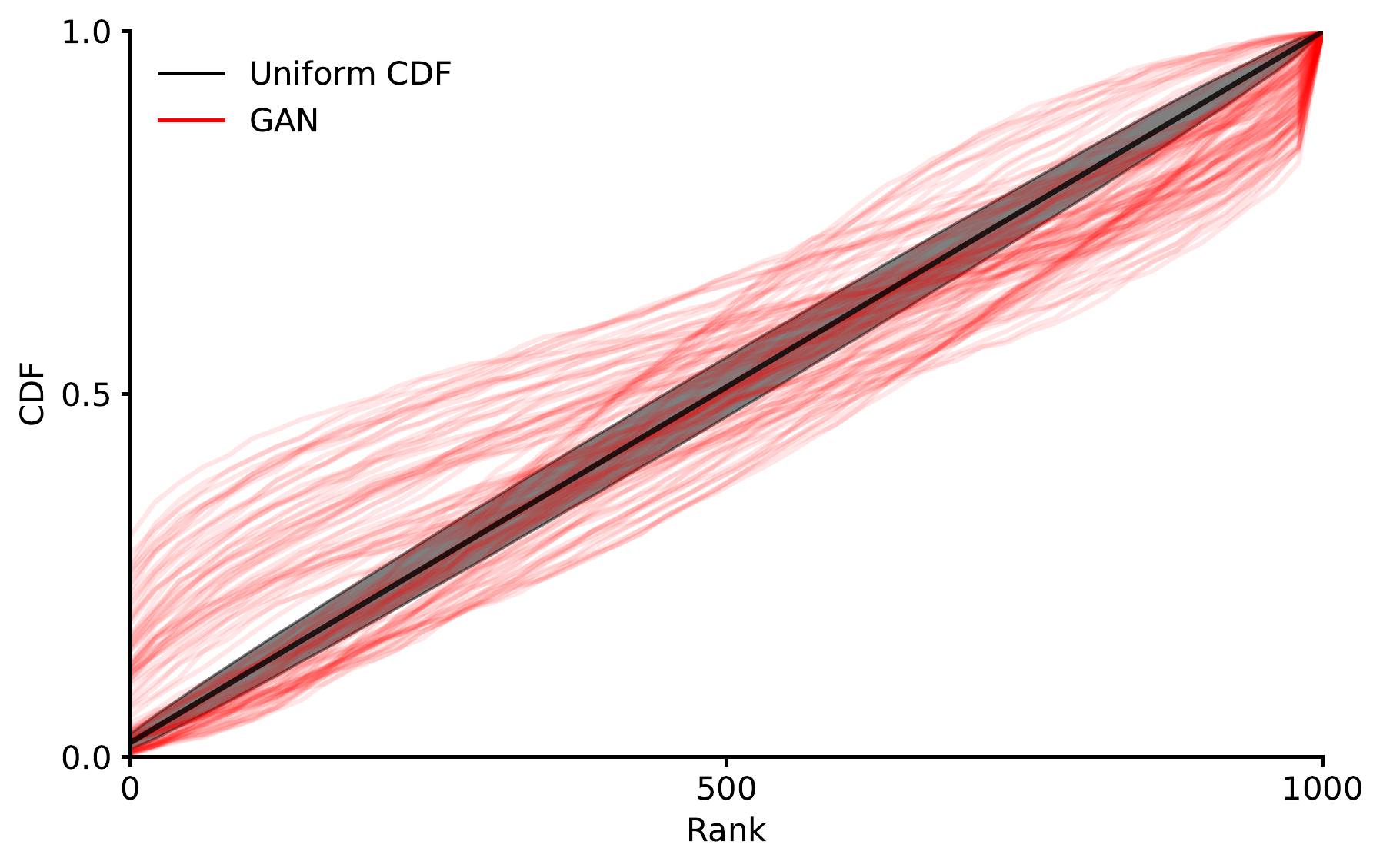} %the gan one, missing
			\end{center}
			\caption{GAN}\label{}
		\end{subfigure}%
		\begin{subfigure}[t]{0.5\textwidth}
			\centering
			\includegraphics[width=\columnwidth]{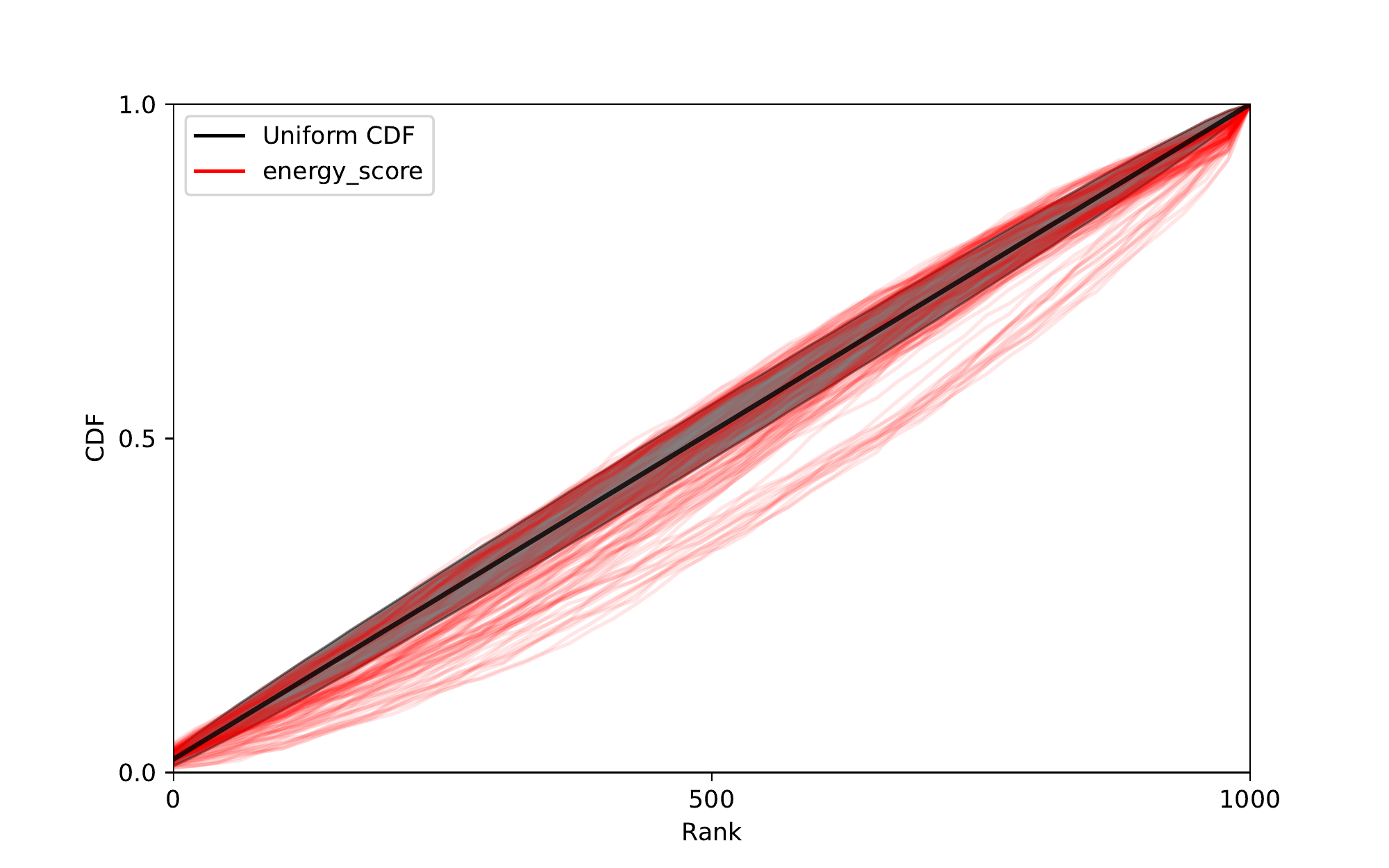}
			\caption{Energy Score, patch size 20 and step 10}\label{}
		\end{subfigure}%\\
		\caption{Shallow Water model: Simulation Based Calibration. Each line corresponds to a single dimension of $ \ddtheta $ and represents the CDF of the rank of the true parameter value with respect to a set of posterior samples. A calibrated posterior implies uniform CDF (diagonal black line, with associated 99\% confidence region for that number of samples in gray).}
		\label{fig:SBC_shallow_water}
	\end{figure}

	\subsection{Noisy Camera model}
	
	Here, we consider $ \ddtheta \in \mathbb{R}^{28\times28}$ to be the images of the EMNIST dataset \citep{cohen2017emnist}, from which the data $ \ddobs  \in \mathbb{R}^{28\times28}$ is generated by applying some blurring (see \citealp{ramesh2022gatsbi} for details). Posterior inference corresponds therefore to Bayesian denoising. In this model, the dimension of parameter space is larger than in typical LFI applications; additionally, the prior is defined implicitly as we can only generate samples from it. This prevents the application of most standard LFI methods. Besides the GAN method, we test here the Energy and Kernel score with $ m=10 $ in three different configurations: 1) on the full parameter space, 2) with patch size 14 and step 7, and 3) with patch size 8 and step 5. Training is done on 800 thousands samples on a NVIDIA Tesla-V100 GPU; additional details are discussed in Appendix~\ref{app:details_camera}. Among the SR methods, those with patch size 8 and step 5 performed better; therefore, we report only results for the Kernel and Energy Score in that configuration in the main body of the paper; results for the other configurations are given in Appendix~\ref{app:res_camera}.

	In Figure~\ref{fig:camera_model_best}, we report posterior mean and standard deviation for a set of observations for the different methods. SR methods lead to cleaner image reconstruction and more meaningful uncertainty quantification.

	% results for camera model
	\begin{figure}
		\centering
		\includegraphics[width=\linewidth]{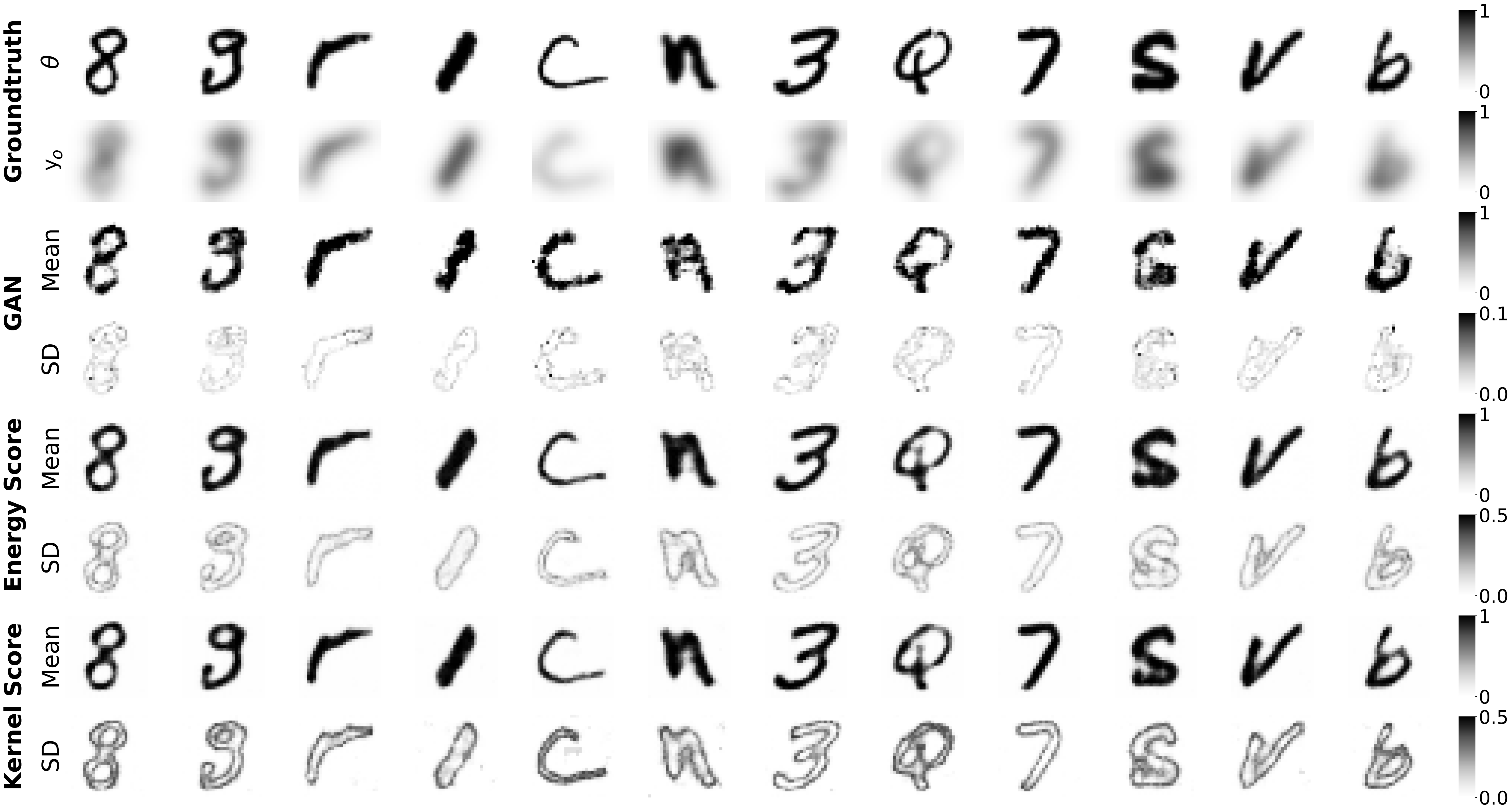}
		\caption{Noisy Camera model: ground truth and posterior inference with different methods, for a set of observations (each observation corresponds to a column). The first two rows represent the ground-truth values of $ \ddtheta $ and the corresponding observation $ \ddobs_o $. The remaining rows represent mean and Standard Deviation (SD) for GAN and Energy and Kernel Score methods with patch size 8 and step 5. Notice how the posterior mean for the SR methods are neater than those obtained with GAN; additionally, the SD is larger close to the boundary of the reconstructed digit (notice the different color scale in the SD for GAN and for the SR methods).}
		\label{fig:camera_model_best}
	\end{figure}
	
	In Table~\ref{tab:camera_model}, we report the performance metrics, runtime and epoch of early stopping of the GAN and SR methods; the latter lead to smaller calibration error, although that is still quite poor in absolute terms. The R$ ^2 $ values here are also poor. We believe these low metric values are due to each pixel only taking a discrete set of values between 0 and 1, with white spaces assigned $ 0 $ and darkest pixels being assigned $ 1 $. The generative network outputs is also bounded in $ (0,1) $ but can never reach 0 or 1 as it is obtained via a continuous transformation from $ \R $. For the calibration error (see Appendix~\ref{app:metric_cal_err}), that means that a credible interval obtained from the generative network cannot contain the extreme values 0 or 1; similarly, the approximate posterior mean can never be smaller than 0 or larger than 1, thus decreasing the R$^2  $ values (see Appendix~\ref{app:metric_r2}).
	Additionally, we report here the un-normalized RMSE, as computing the normalization would lead to infinite values for the pixels in which the true value is 0 for all training samples (see Appendix~\ref{app:metric_RMSE}).

	\begin{table}[htb]
	\caption{Noisy Camera model: performance metrics, runtime and early stopping epoch for GAN and for the Energy and Kernel Score with patch size 8 and step 5. The latter methods achieved better performance with shorter training time. All methods are trained on a single GPU.}% Cal. error and NRMSE: the smaller, the better. R$ ^2 $: the larger, the better.}
	\label{tab:camera_model}
	\begin{center}
\begin{tabular}{lcccrr}
	\hline
	& RMSE $ \downarrow $   & Cal. Err. $ \downarrow $   & R$^2$ $ \uparrow $   &   Runtime (sec) &   Early stopping epoch \\
	\hline
	GAN              & 0.25 $\pm$ 0.19       & 0.50 $\pm$ 0.00            & -23.94 $\pm$ 366.08  &           45398 &                   3600 \\
	Energy   & 0.06 $\pm$ 0.05       & 0.36 $\pm$ 0.12            & -2.14 $\pm$ 55.86    &           22633 &                   4000 \\
	Kernel   & 0.07 $\pm$ 0.05       & 0.36 $\pm$ 0.12            & -10.29 $\pm$ 222.12  &           22545 &                   3200 \\
	\hline
\end{tabular}	\end{center}
\end{table}

	\section{Conclusions}\label{sec:conclusions}
	
	We considered using a generative network to represent posterior distributions for Likelihood-Free Inference, following \cite{ramesh2022gatsbi}, and investigated training it via Scoring Rule minimization rather than in an adversarial setup as was done in \cite{ramesh2022gatsbi}. The Scoring Rule approach is theoretically grounded and does not suffer from training instability and biased gradients, as the adversarial approach does. In simulation studies, and especially on high-dimensional tasks, we found the Scoring Rule approach generally performed better and was substantially cheaper to train. These findings corroborate similar ones reported in \cite{pacchiardi2021probabilistic} in the setting of probabilistic forecasting, making Scoring Rules minimization an appealing method to train generative networks, particularly when uncertainty quantification in the approximate distribution is critical.
	
	In our simulation studies, we only considered the original GAN objective \citep{goodfellow2014generative} in the adversarial setup, which is what \cite{ramesh2022gatsbi} did. We believe more advanced adversarial training would lead to better results; however, in \cite{pacchiardi2021probabilistic}, Scoring Rule minimization was shown to outperform even more advanced adversarial approaches for probabilistic forecasting, while still being cheaper and easier to train. We expect the same holds for Likelihood-Free Inference.
%	than state-of-the-art methods and was easier to train; we expect therefore to have similar results here. 

	\section*{Acknowledgements}
	
	The authors thank Poornima Ramesh for help using the code used to create the results reported in \cite{ramesh2022gatsbi} and for providing additional result files.
	
	LP is supported by the EPSRC and MRC through the OxWaSP CDT programme (EP/L016710/1),
	which also funds the computational resources used to perform this work. RD is funded by EPSRC (grant nos. EP/V025899/1, EP/T017112/1) and NERC (grant no. NE/T00973X/1).
	We thank Geoff Nicholls for valuable feedback and suggestions.

	\bibliographystyle{abbrvnat}
%	\bibliography{../../references.bib}

	\newpage
	\section*{Appendix}
	\appendix

\section{\textit{f}-GAN}\label{app:f-GAN}
The problem in Eq~\eqref{Eq:gan} can be obtained as a relaxation of the following one:
\begin{equation}\label{}
\arg \min_{\phi}  \E_{\Ddobs \sim P}\left[D_{JS}(\Pi(\cdot|\Ddobs) \| Q_\phi(\cdot|\Ddobs))\right],
\end{equation}
where $D_{JS} $ is the Jensen-Shannon divergence. The objective in the above problem is 0 if and only if $ \Pi(\cdot|\ddobs) = Q_\phi(\cdot|\ddobs) $ for each $ \ddobs:p(\ddobs)>0 $. We report here a more general result by considering a class of divergences known as \textit{f}-divergences, to which the Jensen-Shannon one belongs. We follow \cite{nowozin2016f} in doing so\footnote{An analogous procedure allows to obtain a tractable training objective for the 1-Wasserstein distance as well \cite{arjovsky2017wasserstein}}. 

By discarding temporarily dependence on $ \Ddobs $, an \textit{f}-divergence is defined as:
\begin{equation}\label{}
D_f(P||Q_\phi) = \int_{} q_\phi(\ddtheta) f\left(\frac{p(\ddtheta)}{q_\phi(\ddtheta)}\right)d\mu(\ddtheta),
\end{equation}
where $ f:\mathbb{R}_+\to \R $  is a convex, lower-semicontinuous function for which $ f(1)=0 $, and where $ q_\phi $ and $ p$ are densities of $ Q_\phi $ and $ P $ with respect to a base measure $ \mu $. We want now to fix $ \phi $ via: 
\begin{equation}\label{Eq:f-div-min}
\argmin_\phi D_f(P||Q_\phi).
\end{equation}
 Let now $ \operatorname{dom}_f $ denote the domain of $ f $. By exploiting the Fenchel conjugate $ f^*(t) = \sup_{u \in \operatorname{dom}_f } \left\{ut - f(u)\right\}$, \citealp{nowozin2016f} obtain the following variational lower bound: 
\begin{equation}\label{}
D_f(P||Q_\phi)	\ge \sup_{c \in \mathcal{C}} \left(\mathbb{E}_{\ddtheta\sim P}c(\ddtheta )-\mathbb{E}_{\tilde \ddtheta  \sim Q_\phi}f^{*}(c(\tilde \ddtheta ))\right),
%	D_f(P^\star||P^\phi)	\ge \sup_{C \in \mathcal{C}} \left(\mathbb{E}_{\ddobs \sim P^\star}[C(\ddobs)]-\mathbb{E}_{\ddobs \sim P^\phi}\left[f^{*}(C(\ddobs))\right]\right),
\end{equation}
which holds for any set of functions $ \mathcal{C} $ from $\Y $ to $ \operatorname{dom}_{f^*}$.
By considering a parametric set of functions $ \mathcal{C} =\{ c_\psi:\mathcal{Y} \to  \operatorname{dom}_{f^*}, \psi \in \Psi\} $, a surrogate to the problem in Eq.~\eqref{Eq:f-div-min} becomes:
\begin{eqnarray}\label{}
\min_\phi \max_\psi  \left(\mathbb{E}_{\ddtheta\sim P}c_\psi(\ddtheta )-\mathbb{E}_{\tilde \ddtheta  \sim Q_\phi}f^{*}(c_\psi(\tilde \ddtheta ))\right).
\end{eqnarray}
By re-introducing the dependence on $ \Ddobs $, the above generalizes to:
\begin{equation}\label{Eq:f-gan-cond2}
\min_\phi \max_\psi  \E_{ \Ddobs \sim P}\left(\left(\mathbb{E}_{\ddtheta\sim P(\cdot|\Ddobs)}c_\psi(\ddtheta, \Ddobs )-\mathbb{E}_{\tilde \ddtheta  \sim Q_\phi(\cdot|\Ddobs)}f^{*}(c_\psi(\tilde \ddtheta, \Ddobs ))\right)\right),
\end{equation}
where now the function $c_\psi $ also depends on the value of $ \Ddobs $. 
%	If we now observe joint data 
%\begin{equation}\label{Eq:joint_data2}
%(\ddtheta_i, \ddobs_i)\text{, where } \theta_i\sim\Pi \text{ and } \ddobs_i \sim P^\star(\cdot|\ddtheta_i)
%\end{equation} and we want to train $ \phi $ such that $ P^\phi(\cdot|\ddtheta) =P^\star(\cdot|\ddtheta) $ $ \Pi $-almost everywhere, we could attempt solving: 
%\begin{equation}\label{Eq:f-gan-cond2}
%\begin{aligned}
%\min_\phi \max_\psi  \E_{ \ddtheta \sim \Pi}\big(&\mathbb{E}_{\Ddobs \sim P^\star(\cdot|\ddtheta)}c_\psi(\Ddobs; \ddtheta)\\
%&-\mathbb{E}_{\Ddobs \sim P^\phi(\cdot|\ddtheta)}f^{*}(c_\psi(\Ddobs;\ddtheta))\big),
%\end{aligned}
%\end{equation}
%where now $ c_\psi: \mathcal {Y} \times \Theta \to \operatorname{dom}_{f^*}$.

In practice, $ c_\psi $ is parametrized by a Neural Network. To solve the problem in Eq.~\eqref{Eq:f-gan-cond2}, people usually employ alternating optimization over $ \phi $ and $ \psi $ by following stochastic gradients; this technique is called \textit{f}-GAN. With a finite number of steps over $ \psi $, this leads to biased gradient estimates for $ \phi $. In Algorithm \ref{alg:cGAN}, we show a single epoch (i.e., a loop on the full training dataset) of conditional \textit{f}-GAN training; for simplicity, we consider here using a single pair $ (\ddtheta_i, \ddobs_i) $ to estimate the expectations in Eq.~\eqref{Eq:f-gan-cond2} (i.e., the batch size is 1), but using a larger number of samples is possible. Notice how in Algorithm \ref{alg:cGAN} we update the critic once every generator update; however, multiple critic updates can be done at each generator update.  
%	Indeed, the theoretical guarantees for GAN training \cite{goodfellow2014generative} consider the critic to be trained until convergence for each generator update.

% by introducing a variational lower bound depending on an arbitrary function which is then parametrized by the neural network $ c_\psi $. This procedure holds for all so-called \textit{f}-divergences, to which the Jensen-Shannon one belongs (see for instance \citealp{nowozin2016f, pacchiardi2021probabilistic})

\begin{algorithm}
	\caption{Single epoch conditional \textit{f}-GAN training.}
	\label{alg:cGAN}
	\begin{algorithmic}
		\REQUIRE Parametric map $ g_\phi $, critic network $ c_\psi $, learning rates $ \epsilon $, $ \gamma $.
		\FOR{each training pair $( \ddtheta_i, \ddobs_i)$}
		%			\FOR{$t$ from $k$ up to $n-l$}
		\STATE Sample $ \mathbf{\ddz} \sim P_\ddz$ 
		\STATE Obtain $ \tilde \ddtheta^{\phi}_i = g_\phi (\mathbf{\ddz}, \ddobs_i) $
		\STATE Set
		$ \psi \leftarrow \psi + \gamma \cdot \nabla_\psi \Big[ c_\psi(\ddtheta_i, \ddobs_i) -f^*(c_\psi(\tilde \ddtheta^{\phi}_i, \ddobs_i)) \Big] $
		\STATE Set $\phi \leftarrow\phi - \epsilon \cdot \nabla_\phi \Big[ -f^*(c_\psi(\tilde \ddtheta^{\phi}_i, \ddobs_i)) \Big] $
		\ENDFOR
	\end{algorithmic}
\end{algorithm}	

	\section{Unbiased gradient estimates}

	We discuss here how we can get unbiased gradient estimates for the Scoring Rule training objective in Eq.~\eqref{Eq:SR_obj_cond_emp} with respect to the parameters of the generative network $ \phi $. 
	
	In order to do that, we first discuss how to obtain unbiased estimates of the SRs we use across this work. Then, we show how those allow to obtain unbiased gradient estimates. The steps we follow are the same as in \cite{pacchiardi2021probabilistic} for the setting of probabilistic forecasting.

	\subsection{Unbiased scoring rule estimates}\label{app:unbiased_SR}
	
	Assume we have draws $\tilde \ddsim_j\sim P, j=1, \ldots, m $.
	
	\paragraph{Energy Score}
	An unbiased estimate of the energy score can be obtained by unbiasedly estimating the expectations in $\SE^{(\beta)}(P, \ddsim) $ in Eq.~\eqref{Eq:eng_score}:
	\begin{equation}
	\hat S_{\text{E}}^{(\beta)}(P, \ddsim) =\frac{2}{m} \sum_{j=1}^m \left\| \tilde \ddsim_j - \ddsim\right\|_2^\beta - \frac{1}{m(m-1)}\sumjk \left\|\tilde \ddsim_j-\tilde \ddsim_k\right\|_2^\beta.
	\end{equation}
	
	\paragraph{Kernel Score} Similarly to the energy score, we obtain an unbiased estimate of $ S_k(P,\ddsim) $ in Eq.~\eqref{Eq:kernel_score} by:
	\begin{equation}
	\hat S_k(P, \ddsim) = \frac{1}{m(m-1)}\sumjk  k(\tilde \ddsim_j,\tilde \ddsim_k )-\frac{2}{m} \sum_{j=1}^m k(\tilde \ddsim_j,\ddsim).
	\end{equation}
	
	\paragraph{Sum of SRs}	
	When adding multiple SRs, an unbiased gradient of the sum can be obtained by adding unbiased estimates of the two addends.

	\subsection{Unbiased estimate of the training objective}\label{app:unbiased_estimate_obj}

	Recall now we want to solve:	
	\begin{gather}\label{Eq:argmin_problem}
	\hat \phi :=\argmin_\phi J(\phi), \quad  J(\phi) = \frac{1}{n} \sum_{i=1}^n S(Q_\phi(\cdot|\ddobs_i),\ddtheta_i)
	\end{gather}
	In order to do this, we exploit Stochastic Gradient Descent (SGD), which requires unbiased estimates of $ J(\phi) $. 
	Notice how, for all the Scoring Rules used across this work, as well as any weighted sum of those, we can write:
	$ S(P,\ddsim) = \E_{\tilde \Ddsim, \tilde  \Ddsim'\sim P}\left[h(\tilde\Ddsim, \tilde\Ddsim',\ddsim)\right] $ for some function $ h $; namely, the SR is defined through an expectation over (possibly multiple) samples from $ P $. That is the form exploited in Appendix~\ref{app:unbiased_SR} to obtain unbiased SR estimates.
	
	Now, we will use this fact to obtain unbiased estimates for the objective in Eq.~\eqref{Eq:argmin_problem}. 
	%For brevity, let us now denote $ J(\phi ) = 	S_T (Q_\phi_{k+l:T}(\cdot|\ddobs_{1:k+l-1}), \ddobs_{k+l:T}) $, which we can rewrite as (letting	$ N =T-l-k+1 $ for brevity):

	%g_\phi(\cdot;\ddobs_{t-k+1:t})\sharp Q
	\begin{equation}\label{Eq:J2}
	J(\phi)=\frac{1}{n}\sum\limits_{i=1}^{n} \E_{\tilde \ddtheta, \tilde \ddtheta'\sim Q_\phi(\cdot|\ddobs_i)}\left[h(\tilde \ddtheta, \tilde \ddtheta', \ddtheta_i)\right] = \frac{1}{n}\sum\limits_{i=1}^{n} \E_{\mathbf{\Ddz},\mathbf{\Ddz}'\sim P_\ddz}\left[h(g_\phi(\mathbf{\Ddz},\ddobs_i),g_\phi(\mathbf{\Ddz}',\ddobs_i),\ddtheta_i)\right],
	%\sum\limits_{t=0}^{n-1}S(P_\phi^{t,l},\ddobs_{t+l})=.
	\end{equation}
	where we used the fact that $ Q_\phi$ is the distribution induced by a generative network with transformation $ g_\phi$; this is called the reparametrization trick \cite{kingma2013auto}. Now:
	\begin{equation}\label{Eq:J_grad}
	\begin{aligned}
	\nabla_\phi J(\phi)&= \nabla_\phi\frac{1}{n}\sum\limits_{i=1}^{n}  \E_{\mathbf{\Ddz},\mathbf{\Ddz}'\sim P_\ddz}\left[h(g_\phi(\mathbf{\Ddz},\ddobs_i),g_\phi(\mathbf{\Ddz}',\ddobs_i),\ddtheta_i)\right] \\
	&= \frac{1}{n}\sum\limits_{i=1}^{n} \E_{\mathbf{\Ddz},\mathbf{\Ddz}'\sim P_\ddz}\left[\nabla_\phi h(g_\phi(\mathbf{\Ddz},\ddobs_i),g_\phi(\mathbf{\Ddz}',\ddobs_i),\ddtheta_i)\right].
	\end{aligned}
	\end{equation}
	In the latter equality, the exchange between expectation and gradient is not a trivial step, due to the non-differentiability of functions (such as ReLU) used in $ g_\phi $. Luckily, Theorem 5 in \cite{binkowski2018demystifying} proved that to be valid almost surely with respect to a measure on the space $ \Phi $ to which neural network weights $ \phi $ belong, under mild conditions on the NN architecture.
	
	We can now easily obtain an unbiased estimate of the above using samples $ \mathbf z_{i,j}\sim Q, j=1,\ldots,m $, for each $ i\in\{1,\ldots,n\} $.
	%	As mentioned in the main text, we cannot in general compute the expectation above.
	Additionally, Stochastic Gradient Descent 
	usually considers a small batch of training samples at each step, obtained by taking a random subset (or batch) $ \mathcal{B} \subseteq \{ 1, 2, \ldots, n \}$. Therefore, the following unbiased estimate of $ \nabla_\phi J(\phi) $ can be obtained:
	\begin{equation}\label{}
	\widehat {\nabla_\phi J(\phi)}= \frac{1}{|\mathcal{B}|}\sum\limits_{i\in \mathcal{B}}  \frac{1}{m(m-1)} \sumij  \nabla_\phi h(g_\phi(\mathbf z_{i,j};\ddobs_i),g_\phi(\mathbf z_{i,k};\ddobs_i),\ddtheta_i).
	%\sum\limits_{t=0}^{n-1}S(P_\phi^{t,l},\ddobs_{t+l})=.
	\end{equation}
	In practice, the above is obtained by computing the gradient of the following unbiased estimate of $ J(\phi) $ via autodifferentiation libraries (see for instance \citealp{pytorch}):
	\begin{equation}\label{}
	\hat J {(\phi)}= \frac{1}{|\mathcal{B}|}\sum\limits_{i\in \mathcal{B}}  \frac{1}{m(m-1)} \sumij  h(g_\phi(\mathbf z_{i,j};\ddobs_i),g_\phi(\mathbf z_{i,k};\ddobs_i),\ddtheta_i).
		%\sum\limits_{t=0}^{n-1}S(P_\phi^{t,l},\ddobs_{t+l})=.
	\end{equation}

	In Algorithm~\ref{alg:gen-SR-conditional}, we train a generative network for a single epoch using a scoring rule $ S $ for which unbiased estimators can be obtained by using more than one sample from $ Q_\phi $.
	%	, as it is the case for the Kernel Score (see Appendix~\ref{app:unbiased_estimates}). 
	Compare it with the adversarial approach reported in Algorithm~\ref{alg:cGAN}; in the SR approach, multiple samples from the generative networks are required at each step ($ m>1 $), while a unique one is enough for the adversarial approach. Conversely, however, the SR approach does not require an additional critic network and learning rate $ \gamma $ and is simpler and quicker to train (see the results in Sec.~\ref{sec:results} and \citealp{pacchiardi2021probabilistic} for more details).
	As in Algorithm~\ref{alg:cGAN} , we use a single pair $ (\ddtheta_i, \ddobs_i) $ to estimate the gradient.

	\begin{algorithm}
		\caption{Single epoch generative-SR training.}
		\label{alg:gen-SR-conditional}
		\begin{algorithmic}
			\REQUIRE Parametric map $ g_\phi $, SR $ S $, learning rate $ \epsilon $.
			%			\FOR{$t$ from $k$ up to $n-l$}
			\FOR{each training pair $( \ddtheta_i, \ddobs_i)$}
			\STATE Sample {\textbf{multiple}} $ \mathbf z_1, \ldots,\mathbf z_m $ 
			\STATE Obtain $ \tilde \ddtheta_{i,j}^\phi = g_\phi (\mathbf z_j, \ddobs_i)$
			\STATE Obtain unbiased estimate $ \hat S (Q_\phi(\cdot| \ddobs_i), \ddtheta_i)$ from $  \tilde \ddtheta_{i,j}^\phi$%, j=1,\ldots,m $
			\STATE Set $\phi \leftarrow \phi - \epsilon \cdot \nabla_\phi \hat  S (Q_\phi(\cdot| \ddobs_i), \ddtheta_i) $
			\ENDFOR
		\end{algorithmic}		
	\end{algorithm}

	\section{Details on performance measures}\label{app:metric}
	
	We review here the measures of performance used in the empirical studies. We follow \cite{radev2020bayesflow} in defining these measures; we report them here for ease of reference.
	All these metrics are for univariate $ \ddtheta $; when handling multivariate $ \ddtheta $, we therefore compute them on each dimension separately and report the average.

	\subsection{Deterministic performance measures}
	
	We discuss two measures of performance of a deterministic forecast $ \hat \ddtheta_i $ for a realization $ \ddtheta_i $; across our work, we take $ \hat \ddtheta_i  $ to be the mean of the (univariate) probability distribution $  Q_\phi(\cdot|\ddobs_i) $.

	\subsubsection{RMSE}\label{app:metric_RMSE}
	We first introduce the Root Mean-Square Error (RMSE) as:
	$$\operatorname{RMSE} =\sqrt{\frac{1}{n}\sum\limits_{i=1}^{n}\left(\hat \ddtheta_i - \ddtheta_i \right)^{2}},$$
	where we consider here for simplicity $ i=1, \ldots, n $. From the above, we obtain the Normalized RMSE (NRMSE) as:
	$$\operatorname{NRMSE} = \frac{RMSE}{\max_i\{ \ddtheta_i \} - \min_i\{ \ddtheta_i \}}.$$
	$ \operatorname{NRMSE} =0 $ implies $ \hat \ddtheta_i = \ddtheta_i $ for all $ i $'s. NRMSE $ \in[0,1] $ and allows to compare performance over different tasks. Notice however that, when $ \max_i\{ \ddtheta_i \} = \min_i\{ \ddtheta_i\} $, NRMSE diverges; in that case, we consider the un-normalized RMSE.
	
	\subsubsection{Coefficient of determination}\label{app:metric_r2}
	The coefficient of determination $\operatorname{R}^2 $ measures how much of the variance in $ \{\ddtheta_i\}_{i=1}^n $ is explained by $ \{\hat \ddtheta_i\}_{i=1}^n $. Specifically, it is given by:
	
	$$
	\operatorname{R}^{2}=1-\frac{\sum_{i=1}^{n} \left(\ddtheta_i-\hat \ddtheta_i\right)^{2}}{\sum_{i=1}^{n} \left(\ddtheta_i-\bar{\ddtheta} \right)^{2}},
	$$
	where $\bar{\ddtheta} = \frac{1}{n} \sum\limits_{i=1}^{n} \ddtheta_i$. $ \operatorname{R}^{2}\le1 $ and $ \operatorname{R}^{2}=1 \implies \hat \ddtheta_i = \ddtheta_i $ for all $ i $'s.

	\subsection{Calibration measures}
	
	We review here two measures of calibration of a probabilistic forecast. Both measures consider the univariate marginals of the approximate posterior distribution $ Q_\phi(\cdot|\ddobs_i) $; for component $ l $, let us denote that by $ Q_{\phi, l}(\cdot|\ddobs_i) $. 
	
	%univariate marginals of the probabilistic forecast distribution $ \probfor $; for component $ i $, let us denote that by $ P_{\phi,i}(\cdot|\ddobs_{t-k+1:t}) $. 
	
	%\subsubsection{Calibration error}
	
	%We review here a measure of calibration of a probabilistic forecast; this measure considers the univariate marginals of the approximate posterior distribution $ Q_\phi(\cdot|y_i) $; for component $ i $, let us denote that by $ P_{\phi,i}(\cdot|\ddobs_{i-k+1:t}) $.
	%	In defining it, we follow \cite{radev2020bayesflow}.

	\subsubsection{Calibration error}\label{app:metric_cal_err}
	
	The calibration error \cite{radev2020bayesflow} quantifies how well the credible intervals of approximate posterior $\apprpostl $ match the distribution of $ \theta_{i,l} $. Specifically, let $ \alpha(l) $ be the proportion of times the verification $ \theta_{i,l} $ falls into an $ \alpha $-credible interval of $ \apprpostl $, computed over all values of $ i $. If the marginal forecast distribution is perfectly calibrated for component $ l $, $ \alpha(l) =\alpha$ for all values of $ \alpha \in (0,1) $. 
	
	We define therefore the calibration error as the median of $ |\alpha(l) -\alpha |$ over 100 equally spaced values of $ \alpha\in(0,1) $. Therefore, the calibration error is a value between $ 0 $ and 1, where $ 0 $ denotes perfect calibration. 
	
	In practice, the credible intervals of the predictive are estimated using a set of samples from $ \apprpost $.

	\subsubsection{Simulation-Based Calibration (SBC)}\label{app:sbc}
	
	SBC \cite{talts2018validating} tests a self-consistency property of the Bayesian posterior in a posterior approximation. In fact, the Bayesian posterior satisfies the following equality: 
	\begin{equation}
	\label{Eq:SBC}
	%	\begin{aligned}
	\pi(\ddtheta) =\int p(\ddtheta, \tilde{\ddtheta}, \tilde{\ddobs}) d \tilde{\ddobs} d \tilde{\ddtheta} 
	=\int p(\ddtheta, \tilde{\ddobs} \mid \tilde{\ddtheta}) \pi(\tilde{\ddtheta}) d \tilde{\ddobs} d \tilde{\ddtheta} 
	=\int \pi(\ddtheta \mid \tilde{\ddobs}) p(\tilde{\ddobs} \mid \tilde{\ddtheta}) \pi(\tilde{\ddtheta}) d \tilde{\ddobs} d \tilde{\ddtheta}
	%	\end{aligned}
	\end{equation}
	in practice, this means that, if you sample from the prior $ \tilde \ddtheta \sim \pi$, use that to generate a sample from the likelihood $ \tilde \ddobs \sim p(\cdot|\ddtheta) $ and use the latter in turn to generate a posterior sample $ \ddtheta \sim \pi(\cdot|\tilde \ddobs) $, $ \ddtheta $ is distributed according to the prior $ \pi(\ddtheta) $. If you repeat the same procedure by sampling $ \ddtheta $ from an \textit{approximate} posterior, say $ \ddtheta \sim Q_\phi(\cdot|\tilde \ddobs) $, then $ \ddtheta \sim \pi $ is a necessary condition for $ q_\phi(\cdot|\ddobs) =\pi(\cdot|\ddobs)$, i.e. for the approximate posterior to be exact. Notice however how this is \textit{not} a sufficient condition: the equality can be satisfied even if $ q_\phi(\cdot|\ddobs) $ is different from the posterior (it is in fact trivially satisfied $ q_\phi(\cdot|\ddobs) = \pi $, i.e. when the approximate posterior corresponds to the prior).
	
	A way to empirically test the above property involves, for a given prior sample $ \tilde \ddtheta $, drawing from the likelihood multiple times $ \ddobs_i \sim p(\cdot|\tilde \ddtheta), i=1, \ldots, N $ and, for each of these, obtaining a single approximate posterior sample $ \ddtheta_i \sim q_\phi(\cdot|\ddobs_i)$. Given these, you compute the rank of $ \tilde \ddtheta $: $r=\sum_{i=1}^{N} \mathbf{1}_{\left[\ddtheta_i<\tilde{\ddtheta}\right]}$ (this only makes sense if $ \ddtheta $ is univariate; otherwise, you compute the rank independently for each dimension of $ \ddtheta $). If $ \ddtheta_i $'s were effectively distributed from the prior, $ r $ is a uniform random variable on $ \{1,2,\ldots,N\} $. Therefore, by repeating this procedure for different prior samples $ \tilde \ddtheta $ and visualizing the distribution of the resulting $ r $'s (for instance via an histogram or by plotting the CDF) gives an indication of whether an equivalence such as Eq.~\eqref{Eq:SBC} is satisfied for $ q_\phi $. See Algorithm 2 in \cite{radev2020bayesflow} for a precise description of this procedure, which goes under the name of Simulation-Based Calibration. This is closely related to the concept of probabilistic calibration and rank histogram in the framework of probabilistic forecasting \citep{gneiting2007probabilistic}.

	\section{Experimental details}\label{app:training_details}
	
	Precise configuration details can be found in the \href{https://github.com/LoryPack/LFI_gen_networks_SRs}{code accompanying the paper}. 	
	\subsection{Benchmark models} 
	Except for the details reported in the main body of the paper, the training configuration for the two benchmark models is the same as in \cite{ramesh2022gatsbi}; of course, some hyperparameter values for the GAN training routine do not apply to the SR one (for instance, all the hyperparameters related to the discriminator).

	\subsection{Shallow Water Model}\label{app:details_shallow_water}
	We train all methods for at most 40k epochs on 100k training samples. For the SR method, we tried both $ m=3 $ and $ m=10  $, with the latter resulting in improved performance; all results reported across the paper refer therefore to $ m=10 $.
	
	GAN used a batch size of 125 (as in \citealp{ramesh2022gatsbi}), while the SR methods used a batch size of 60 (otherwise GPU memory overflow occurs). 
	
	Recall that the parameters $ \ddtheta\in\mathbb{R}^{100} $ are disposed along a 1D uniform grid. 
	When using the patched SR configuration, we consider patches of size \texttt{patch\_size} disposed at a distance \texttt{patch\_step} one from the other. The number of patches is therefore $$ \texttt{n\_patches} = (100 - \texttt{patch\_size}) / \texttt{patch\_step} + 1. $$
	We used therefore the following patched SR configurations on the 1D grid:
	\begin{enumerate}
		\item \texttt{patch\_size} $ =10 $ and \texttt{patch\_step}$ =5 $, which results in $ \texttt{n\_patches} =19 $.
		\item \texttt{patch\_size} $ =20 $ and \texttt{patch\_step}$ =10 $, which results in $ \texttt{n\_patches}=9 $.
	\end{enumerate}
	The patched SR is added to the overall score over the full parameter space.% \textbf{WEIGHT}
	
	The training time (per epoch) is roughly constant in the un-patched and the two different patched configurations.

	\subsection{Camera Model}\label{app:details_camera}
	We train all methods for at most 10k epochs on 800k training samples. For the SR method, we tried both $ m=3 $ and $ m=10  $, with the latter resulting in improved performance. 
	
	Both SR and GAN methods used a batch size of $ 800 $ as in \cite{ramesh2022gatsbi}.

	Here, the parameters $ \ddtheta$ is on a $ 28\times 28 $ square grid.
	When using the patched SR configuration, we consider patches of size \texttt{patch\_size}$ \times $\texttt{patch\_size} disposed at a distance \texttt{patch\_step} one from the other in both spatial dimensions. the number of patches is obtained as 
	$$ \texttt{n\_patches} = [(28 - \texttt{patch\_size}) / \texttt{patch\_step} + 1]^2.$$
	We used therefore the following patched SR configurations on the 2D grid:
	\begin{enumerate}
		\item \texttt{patch\_size} $ =14 $ and \texttt{patch\_step} $ =7 $, which results in $ \texttt{n\_patches}=9 $.
		%	$ 3\times3=9 $ patches.
		\item \texttt{patch\_size}$ =8 $ and \texttt{patch\_step} $ =5 $, which results in $ \texttt{n\_patches}=25 $.
		%	$ 5\times5=25 $ patches.
	\end{enumerate}
	The patched SR is added to the overall score over the full parameter space.% \textbf{WEIGHT}
	
	The training time (per epoch) is roughly constant in the un-patched and the two different patched configurations.

	\section{Additional experimental results}

	%posterior samples
	\FloatBarrier
	\subsection{SLCP}\label{app:res_slcp}
	\FloatBarrier
	
\begin{figure}
		\centering
		\includegraphics[width=0.8\linewidth]{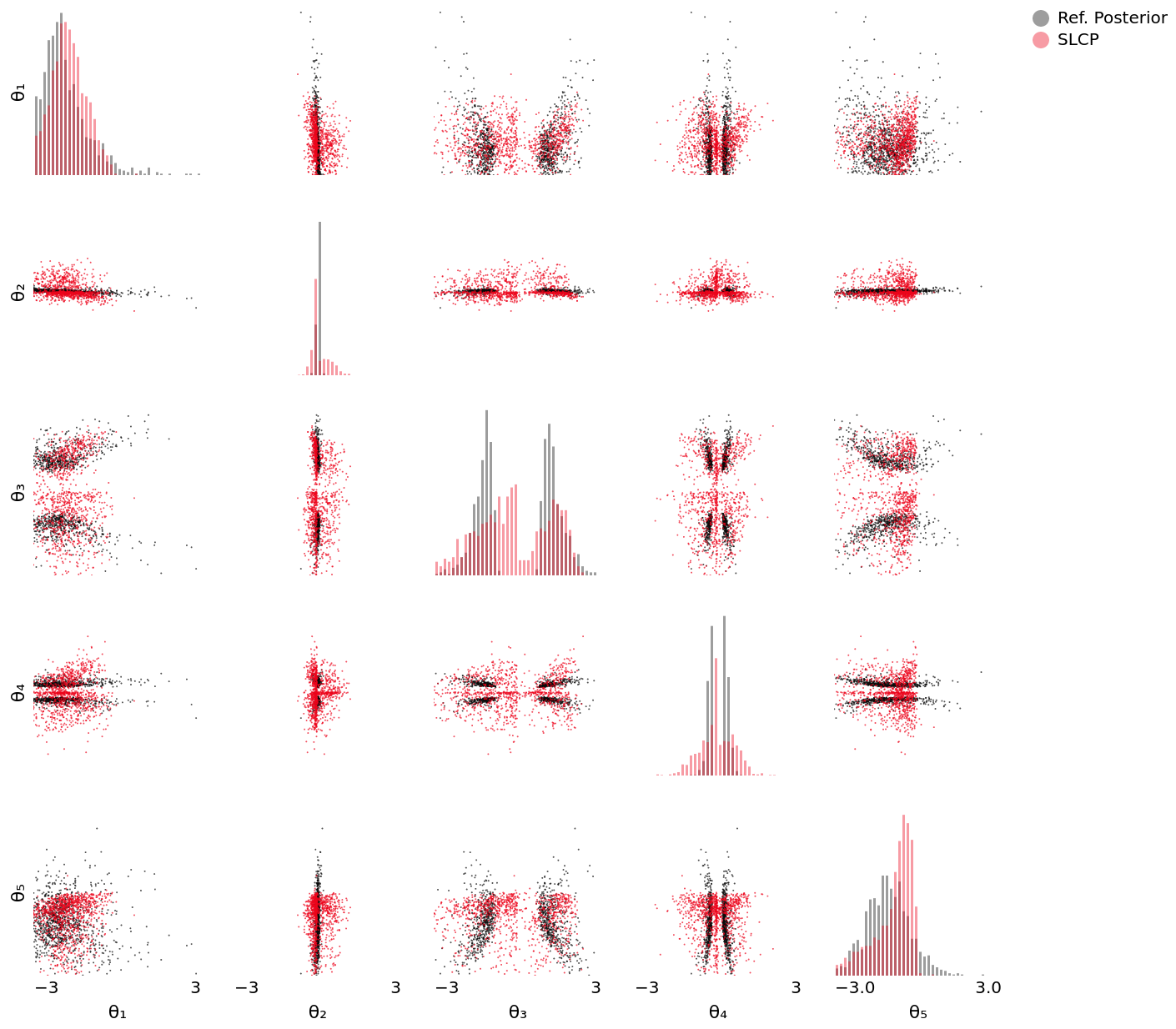}
		\caption{SLCP: posterior samples for Energy Score trained with $ m = 20$ and reference posterior samples. Diagonal panels represent univariate marginals, while off-diagonals represent bivariate marginals. A similar plot for GAN can be found in the supplementary material in \cite{ramesh2022gatsbi}.}
		\label{fig:SLCP_post}
	\end{figure}

	In Figure~\ref{fig:SLCP_post}, we report posterior samples obtained with the Energy Score with $ m=20 $ and compare them with  samples from the reference posterior. In Figure~\ref{fig:SBC_slcp}, we report Simulation-Based Calibration results (see Appendix~\ref{app:sbc}): for each dimension of $ \ddtheta $, each histogram represents the distribution of the rank of the true parameter value in a set of samples from the approximate posterior. We show that for GAN and for the Energy Score with $ m=20 $.
	
	%SBC HIST PLOT: 
	\begin{figure}[ht]
		\centering
		\begin{subfigure}[t]{\textwidth}
			\begin{center}
				\includegraphics[width=\columnwidth]{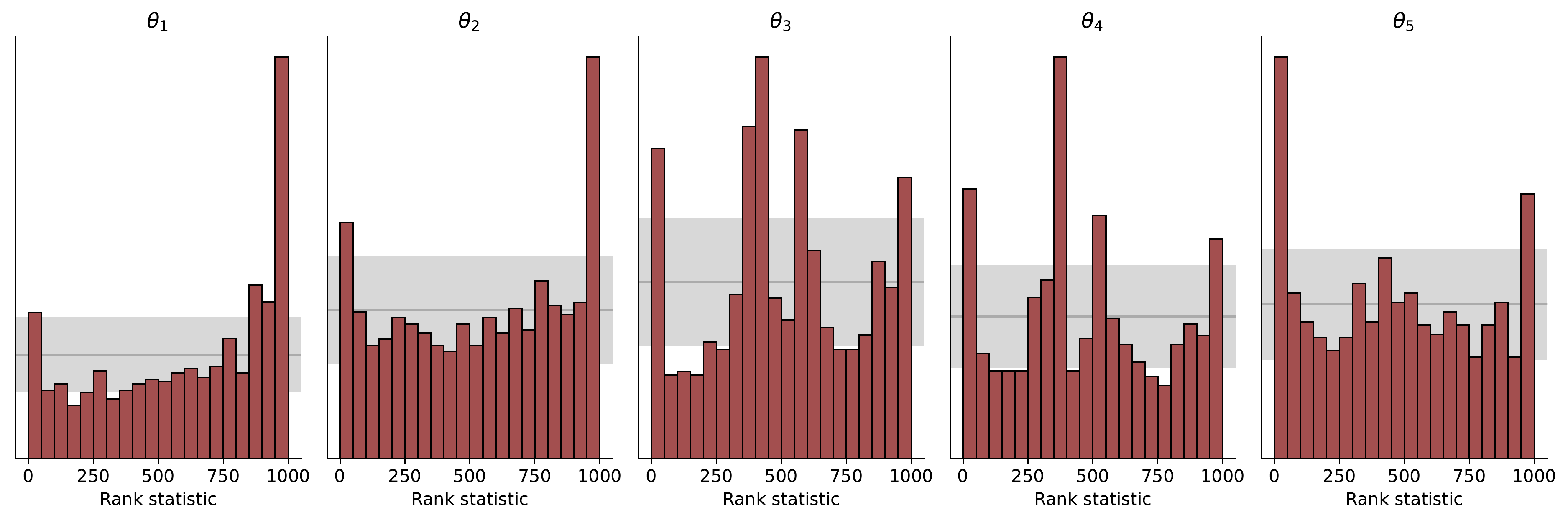}
			\end{center}
			\caption{GAN}\label{}
		\end{subfigure}\\
		\begin{subfigure}[t]{\textwidth}
			\centering
			\includegraphics[width=\columnwidth]{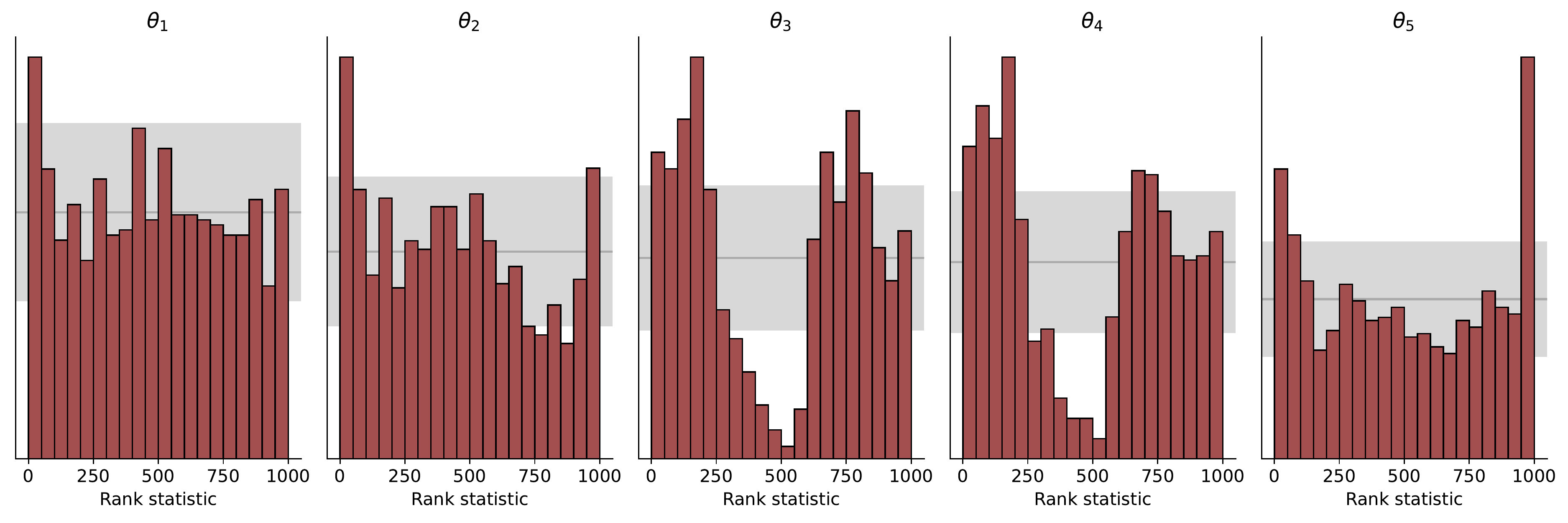}
			\caption{Energy Score, $ m=20 $}\label{}
		\end{subfigure}%\\
		\caption{SLCP: Simulation-Based Calibration results represented as rank histograms; for each dimension of $ \ddtheta $, each histogram represents the distribution of the rank of the true parameter value in a set of samples from the approximate posterior. If the approximate posterior is calibrated, histogram bars should be in the grey region with 99\% probability.}
		\label{fig:SBC_slcp}
	\end{figure}

	% now add all the detailed tables: 
	Tables~\ref{tab:SLCP_C2ST}, \ref{tab:SLCP_RMSE}, \ref{tab:SLCP_cal_error}, \ref{tab:SLCP_r2}, \ref{tab:SLCP_runtime} and \ref{tab:SLCP_es} report the different performance metrics, the runtime and the early stopping epoch for all methods (columns) and all number of  training samples (rows); for Energy and Kernel Score, the number in the column header denotes the number of draws from the generative network during training for each $ \ddobs_i $ in the training batch.
	
	\begin{table}[htb]
		\caption{SLCP: classification-based two-sample test (C2ST).}
		\label{tab:SLCP_C2ST}
		\begin{center}
			\begin{adjustbox}{max width=\textwidth}
				\begin{tabular}{rlllllllll}
\hline
        & GAN             & Energy 3        & Energy 5        & Energy 10       & Energy 20       & Kernel 3        & Kernel 5        & Kernel 10       & Kernel 20       \\
\hline
   1000 & 0.97 $\pm$ 0.02 & 0.99 $\pm$ 0.01 & 0.99 $\pm$ 0.01 & 0.99 $\pm$ 0.00 & 0.99 $\pm$ 0.01 & 1.00 $\pm$ 0.01 & 0.99 $\pm$ 0.01 & 0.99 $\pm$ 0.01 & 0.99 $\pm$ 0.01 \\
  10000 & 0.94 $\pm$ 0.03 & 0.98 $\pm$ 0.01 & 0.97 $\pm$ 0.01 & 0.98 $\pm$ 0.01 & 0.98 $\pm$ 0.01 & 0.99 $\pm$ 0.01 & 0.99 $\pm$ 0.01 & 0.99 $\pm$ 0.01 & 0.99 $\pm$ 0.01 \\
 100000 & 0.92 $\pm$ 0.03 & 0.97 $\pm$ 0.01 & 0.97 $\pm$ 0.02 & 0.96 $\pm$ 0.02 & 0.95 $\pm$ 0.02 & 0.98 $\pm$ 0.01 & 0.98 $\pm$ 0.01 & 0.98 $\pm$ 0.01 & 0.98 $\pm$ 0.01 \\
\hline
\end{tabular}
			\end{adjustbox}		
		\end{center}
	\end{table}

	\begin{table}[htb]
		\caption{SLCP: NRMSE.}% Cal. error and NRMSE: the smaller, the better. R$ ^2 $: the larger, the better.}
		\label{tab:SLCP_RMSE}
		\begin{center}
			\begin{adjustbox}{max width=\textwidth}
				\begin{tabular}{rlllllllll}
\hline
        & GAN             & Energy 3        & Energy 5        & Energy 10       & Energy 20       & Kernel 3        & Kernel 5        & Kernel 10       & Kernel 20       \\
\hline
   1000 & 0.24 $\pm$ 0.05 & 0.25 $\pm$ 0.05 & 0.25 $\pm$ 0.05 & 0.25 $\pm$ 0.05 & 0.25 $\pm$ 0.06 & 0.25 $\pm$ 0.05 & 0.25 $\pm$ 0.05 & 0.25 $\pm$ 0.05 & 0.25 $\pm$ 0.05 \\
  10000 & 0.23 $\pm$ 0.05 & 0.23 $\pm$ 0.05 & 0.23 $\pm$ 0.05 & 0.23 $\pm$ 0.05 & 0.23 $\pm$ 0.05 & 0.23 $\pm$ 0.05 & 0.23 $\pm$ 0.05 & 0.23 $\pm$ 0.05 & 0.23 $\pm$ 0.05 \\
 100000 & 0.23 $\pm$ 0.05 & 0.22 $\pm$ 0.05 & 0.22 $\pm$ 0.06 & 0.22 $\pm$ 0.06 & 0.22 $\pm$ 0.06 & 0.22 $\pm$ 0.06 & 0.22 $\pm$ 0.06 & 0.22 $\pm$ 0.05 & 0.22 $\pm$ 0.06 \\
\hline
\end{tabular}
			\end{adjustbox}		
		\end{center}
	\end{table}

	\begin{table}[htb]
		\caption{SLCP: calibration error.}% Cal. error and NRMSE: the smaller, the better. R$ ^2 $: the larger, the better.}
		\label{tab:SLCP_cal_error}
		\begin{center}
			\begin{adjustbox}{max width=\textwidth}
				\begin{tabular}{rlllllllll}
\hline
        & GAN             & Energy 3        & Energy 5        & Energy 10       & Energy 20       & Kernel 3        & Kernel 5        & Kernel 10       & Kernel 20       \\
\hline
   1000 & 0.13 $\pm$ 0.05 & 0.19 $\pm$ 0.07 & 0.20 $\pm$ 0.05 & 0.20 $\pm$ 0.05 & 0.22 $\pm$ 0.07 & 0.24 $\pm$ 0.09 & 0.23 $\pm$ 0.10 & 0.24 $\pm$ 0.08 & 0.24 $\pm$ 0.08 \\
  10000 & 0.08 $\pm$ 0.03 & 0.11 $\pm$ 0.05 & 0.10 $\pm$ 0.05 & 0.12 $\pm$ 0.07 & 0.10 $\pm$ 0.07 & 0.15 $\pm$ 0.10 & 0.13 $\pm$ 0.09 & 0.14 $\pm$ 0.10 & 0.16 $\pm$ 0.10 \\
 100000 & 0.06 $\pm$ 0.03 & 0.08 $\pm$ 0.07 & 0.08 $\pm$ 0.04 & 0.07 $\pm$ 0.05 & 0.07 $\pm$ 0.04 & 0.13 $\pm$ 0.11 & 0.13 $\pm$ 0.10 & 0.12 $\pm$ 0.08 & 0.13 $\pm$ 0.10 \\
\hline
\end{tabular}
			\end{adjustbox}		
		\end{center}
	\end{table}

	\begin{table}[htb]
		\caption{SLCP: R$^2  $.}% Cal. error and NRMSE: the smaller, the better. R$ ^2 $: the larger, the better.}
		\label{tab:SLCP_r2}
		\begin{center}
			\begin{adjustbox}{max width=\textwidth}
				\begin{tabular}{rlllllllll}
\hline
        & GAN             & Energy 3        & Energy 5        & Energy 10       & Energy 20       & Kernel 3        & Kernel 5        & Kernel 10       & Kernel 20       \\
\hline
   1000 & 0.25 $\pm$ 0.29 & 0.24 $\pm$ 0.30 & 0.22 $\pm$ 0.30 & 0.25 $\pm$ 0.31 & 0.18 $\pm$ 0.35 & 0.22 $\pm$ 0.30 & 0.22 $\pm$ 0.30 & 0.24 $\pm$ 0.30 & 0.23 $\pm$ 0.31 \\
  10000 & 0.35 $\pm$ 0.30 & 0.35 $\pm$ 0.30 & 0.35 $\pm$ 0.30 & 0.35 $\pm$ 0.30 & 0.34 $\pm$ 0.31 & 0.35 $\pm$ 0.29 & 0.35 $\pm$ 0.30 & 0.34 $\pm$ 0.30 & 0.34 $\pm$ 0.30 \\
 100000 & 0.35 $\pm$ 0.30 & 0.36 $\pm$ 0.30 & 0.37 $\pm$ 0.30 & 0.38 $\pm$ 0.32 & 0.38 $\pm$ 0.32 & 0.37 $\pm$ 0.31 & 0.36 $\pm$ 0.31 & 0.36 $\pm$ 0.30 & 0.37 $\pm$ 0.31 \\
\hline
\end{tabular}
			\end{adjustbox}		
		\end{center}
	\end{table}

	\begin{table}[htb]
		\caption{SLCP: runtime in seconds; recall that GAN was trained on GPU while the SR methods were trained on a single CPU.}% Cal. error and NRMSE: the smaller, the better. R$ ^2 $: the larger, the better.}
		\label{tab:SLCP_runtime}
		\begin{center}
			\begin{adjustbox}{max width=\textwidth}
				\begin{tabular}{rrrrrrrrrr}
\hline
        &   GAN &   Energy 3 &   Energy 5 &   Energy 10 &   Energy 20 &   Kernel 3 &   Kernel 5 &   Kernel 10 &   Kernel 20 \\
\hline
   1000 &  4796 &        654 &        692 &         620 &         885 &        515 &        531 &         682 &        1330 \\
  10000 &  9671 &        651 &        658 &         639 &         720 &        636 &        658 &         655 &         697 \\
 100000 & 30963 &       1060 &       1160 &        1305 &        1645 &       1245 &       1044 &        1057 &        1210 \\
\hline
\end{tabular}
			\end{adjustbox}		
		\end{center}
	\end{table}
	
	\begin{table}[htb]
		\caption{SLCP: epoch at which early stopping occurred; the max number of training epochs was 20000.}% Cal. error and NRMSE: the smaller, the better. R$ ^2 $: the larger, the better.}
		\label{tab:SLCP_es}
		\begin{center}
			\begin{adjustbox}{max width=\textwidth}
				\begin{tabular}{rrrrrrrrrr}
\hline
        &   GAN &   Energy 3 &   Energy 5 &   Energy 10 &   Energy 20 &   Kernel 3 &   Kernel 5 &   Kernel 10 &   Kernel 20 \\
\hline
   1000 & 20000 &       1000 &       1000 &        1000 &        1100 &       1100 &       1000 &        1000 &        1000 \\
  10000 & 20000 &       1100 &       1000 &        1100 &        1100 &       1100 &       1100 &        1000 &        1000 \\
 100000 & 20000 &       1000 &       1200 &        1500 &        2100 &       1600 &       1100 &        1000 &        1200 \\
\hline
\end{tabular}
			\end{adjustbox}		
		\end{center}
	\end{table}

	\FloatBarrier
	\subsection{Two Moons}\label{app:res_two_moons}
	\FloatBarrier

\begin{figure}
		\centering
		\includegraphics[width=0.5\linewidth]{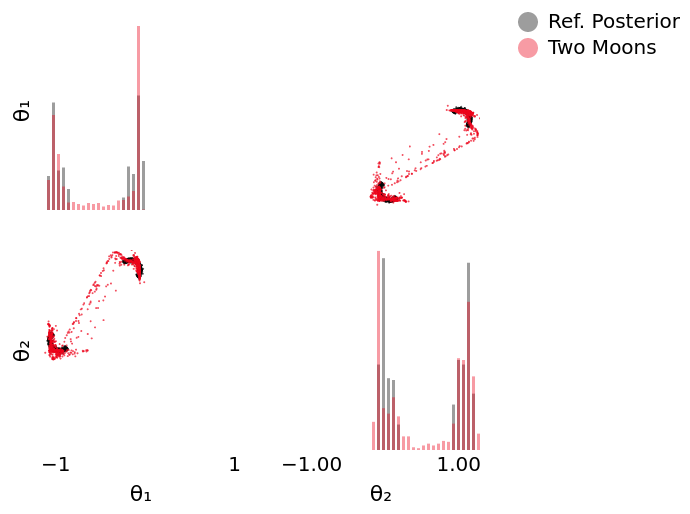}
		\caption{Two Moons: posterior samples for Energy Score trained with $ m = 20$ and reference posterior samples. Diagonal panels represent univariate marginals, while off-diagonals represent bivariate marginals. A similar plot for GAN can be found in the supplementary material in \cite{ramesh2022gatsbi}.}
		\label{fig:two_moons_post}
	\end{figure}

		In Figure~\ref{fig:two_moons_post}, we report posterior samples obtained with the Energy Score with $ m=20 $ and compare them with  samples from the reference posterior. In Figure~\ref{fig:SBC_two_moons}, we report Simulation-Based Calibration results (see Appendix~\ref{app:sbc}): for each dimension of $ \ddtheta $, each histogram represents the distribution of the rank of the true parameter value in a set of samples from the approximate posterior. We show that for GAN and for the Energy Score with $ m=20 $.

	%posterior samples
	% should I add what I get with the GAN? But that has been already put in the original GATSBI draft

	%SBC HIST PLOT: 
	\begin{figure}[ht]
		\centering
		\begin{subfigure}[t]{\textwidth}
			\begin{center}
				\includegraphics[width=0.7\columnwidth]{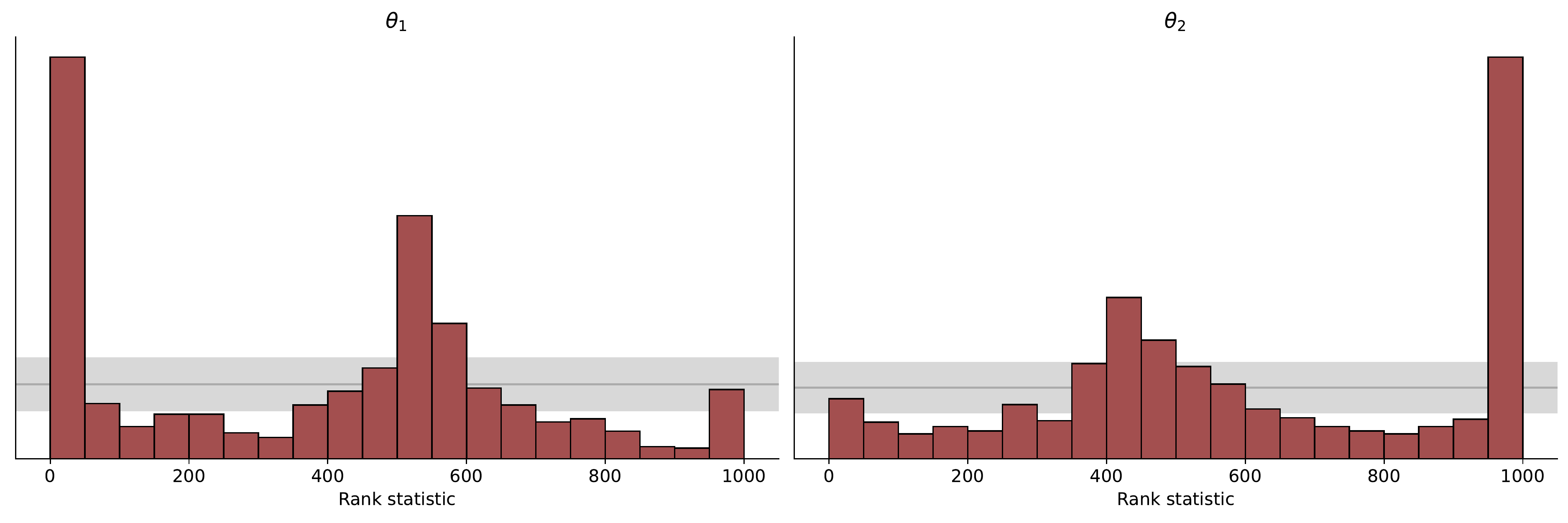}
			\end{center}
			\caption{GAN}\label{}
		\end{subfigure}\\
		\begin{subfigure}[t]{\textwidth}
			\centering
			\includegraphics[width=0.7\columnwidth]{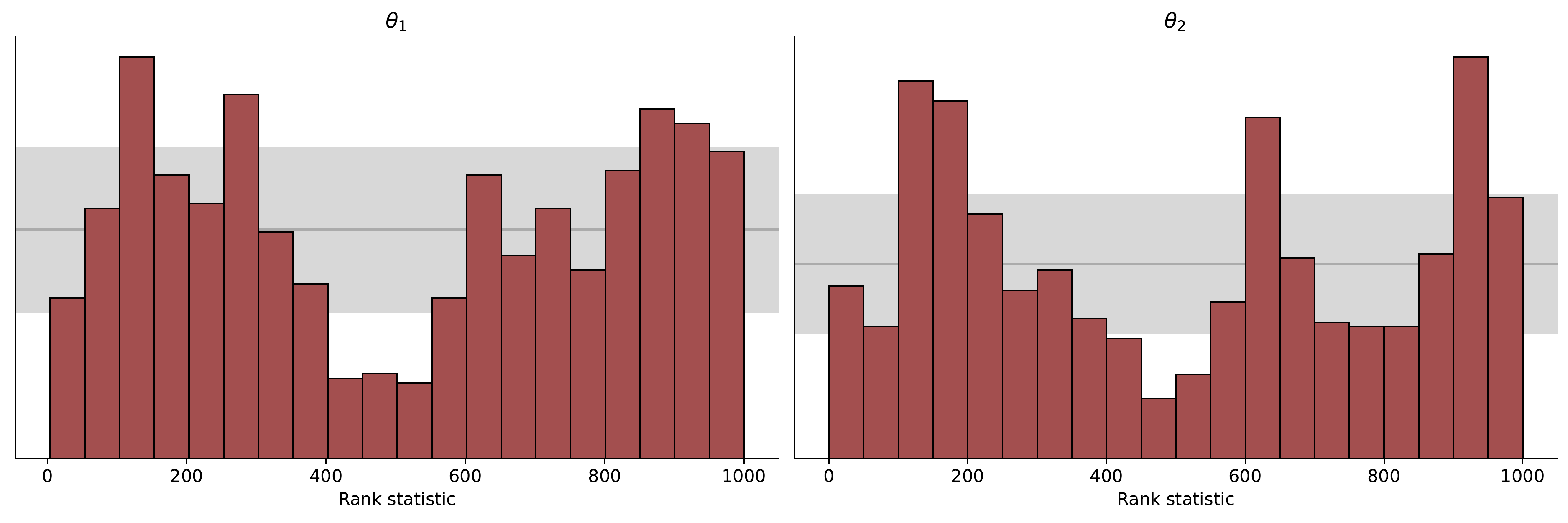}
			\caption{Energy Score, $ m=20 $}\label{}
		\end{subfigure}%\\
		\caption{Two Moons: Simulation-Based Calibration results represented as rank histograms; for each dimension of $ \ddtheta $, each histogram represents the distribution of the rank of the true parameter value in a set of samples from the approximate posterior. If the approximate posterior is calibrated, histogram bars should be in the grey region with 99\% probability.}
		\label{fig:SBC_two_moons}
	\end{figure}

	% now add all the detailed tables: 

	Tables~\ref{tab:Two_Moons_C2ST}, \ref{tab:Two_Moons_RMSE}, \ref{tab:Two_Moons_cal_error}, \ref{tab:Two_Moons_r2}, \ref{tab:Two_Moons_runtime} and \ref{tab:Two_Moons_es} report the different performance metrics, the runtime and the early stopping epoch for all methods (columns) and all number of  training samples (rows); for Energy and Kernel Score, the number in the column header denotes the number of draws from the generative network during training for each $ \ddobs_i $ in the training batch.

	\begin{table}[htb]
		\caption{Two Moonsclassification-based two-sample test (C2ST).}
		\label{tab:Two_Moons_C2ST}
		\begin{center}
			\begin{adjustbox}{max width=\textwidth}
				\begin{tabular}{rlllllllll}
\hline
        & GAN             & Energy 3        & Energy 5        & Energy 10       & Energy 20       & Kernel 3        & Kernel 5        & Kernel 10       & Kernel 20       \\
\hline
   1000 & 0.85 $\pm$ 0.05 & 0.85 $\pm$ 0.06 & 0.87 $\pm$ 0.05 & 0.85 $\pm$ 0.03 & 0.85 $\pm$ 0.04 & 0.94 $\pm$ 0.03 & 0.94 $\pm$ 0.02 & 0.93 $\pm$ 0.03 & 0.96 $\pm$ 0.02 \\
  10000 & 0.81 $\pm$ 0.03 & 0.79 $\pm$ 0.04 & 0.76 $\pm$ 0.05 & 0.76 $\pm$ 0.04 & 0.74 $\pm$ 0.07 & 0.92 $\pm$ 0.03 & 0.93 $\pm$ 0.01 & 0.91 $\pm$ 0.03 & 0.93 $\pm$ 0.01 \\
 100000 & 0.82 $\pm$ 0.07 & 0.79 $\pm$ 0.03 & 0.74 $\pm$ 0.06 & 0.73 $\pm$ 0.05 & 0.73 $\pm$ 0.04 & 0.90 $\pm$ 0.04 & 0.92 $\pm$ 0.03 & 0.90 $\pm$ 0.02 & 0.92 $\pm$ 0.02 \\
\hline
\end{tabular}
			\end{adjustbox}		
		\end{center}
	\end{table}

	\begin{table}[htb]
		\caption{Two Moons: NRMSE.}% Cal. error and NRMSE: the smaller, the better. R$ ^2 $: the larger, the better.}
		\label{tab:Two_Moons_RMSE}
		\begin{center}
			\begin{adjustbox}{max width=\textwidth}
				\begin{tabular}{rlllllllll}
\hline
        & GAN             & Energy 3        & Energy 5        & Energy 10       & Energy 20       & Kernel 3        & Kernel 5        & Kernel 10       & Kernel 20       \\
\hline
   1000 & 0.20 $\pm$ 0.00 & 0.20 $\pm$ 0.00 & 0.20 $\pm$ 0.00 & 0.20 $\pm$ 0.00 & 0.20 $\pm$ 0.00 & 0.21 $\pm$ 0.00 & 0.21 $\pm$ 0.00 & 0.21 $\pm$ 0.00 & 0.20 $\pm$ 0.00 \\
  10000 & 0.20 $\pm$ 0.00 & 0.20 $\pm$ 0.00 & 0.20 $\pm$ 0.00 & 0.20 $\pm$ 0.00 & 0.20 $\pm$ 0.00 & 0.20 $\pm$ 0.00 & 0.20 $\pm$ 0.00 & 0.20 $\pm$ 0.00 & 0.20 $\pm$ 0.00 \\
 100000 & 0.20 $\pm$ 0.00 & 0.20 $\pm$ 0.00 & 0.20 $\pm$ 0.00 & 0.20 $\pm$ 0.00 & 0.20 $\pm$ 0.00 & 0.20 $\pm$ 0.00 & 0.20 $\pm$ 0.00 & 0.20 $\pm$ 0.00 & 0.20 $\pm$ 0.00 \\
\hline
\end{tabular}
			\end{adjustbox}		
		\end{center}
	\end{table}

	\begin{table}[htb]
		\caption{Two Moons: calibration error.}% Cal. error and NRMSE: the smaller, the better. R$ ^2 $: the larger, the better.}
		\label{tab:Two_Moons_cal_error}
		\begin{center}
			\begin{adjustbox}{max width=\textwidth}
				\begin{tabular}{rlllllllll}
\hline
        & GAN             & Energy 3        & Energy 5        & Energy 10       & Energy 20       & Kernel 3        & Kernel 5        & Kernel 10       & Kernel 20       \\
\hline
   1000 & 0.07 $\pm$ 0.01 & 0.05 $\pm$ 0.01 & 0.09 $\pm$ 0.02 & 0.07 $\pm$ 0.01 & 0.06 $\pm$ 0.00 & 0.08 $\pm$ 0.01 & 0.11 $\pm$ 0.00 & 0.14 $\pm$ 0.02 & 0.12 $\pm$ 0.01 \\
  10000 & 0.06 $\pm$ 0.01 & 0.04 $\pm$ 0.02 & 0.03 $\pm$ 0.01 & 0.04 $\pm$ 0.03 & 0.03 $\pm$ 0.01 & 0.04 $\pm$ 0.00 & 0.03 $\pm$ 0.01 & 0.03 $\pm$ 0.02 & 0.03 $\pm$ 0.01 \\
 100000 & 0.07 $\pm$ 0.02 & 0.04 $\pm$ 0.01 & 0.03 $\pm$ 0.00 & 0.04 $\pm$ 0.02 & 0.03 $\pm$ 0.00 & 0.04 $\pm$ 0.00 & 0.03 $\pm$ 0.01 & 0.06 $\pm$ 0.01 & 0.03 $\pm$ 0.01 \\
\hline
\end{tabular}
			\end{adjustbox}		
		\end{center}
	\end{table}

	\begin{table}[htb]
		\caption{Two Moons: R$^2  $.}% Cal. error and NRMSE: the smaller, 
		\label{tab:Two_Moons_r2}
		\begin{center}
			\begin{adjustbox}{max width=\textwidth}
				\begin{tabular}{rlllllllll}
\hline
        & GAN             & Energy 3        & Energy 5        & Energy 10       & Energy 20       & Kernel 3        & Kernel 5        & Kernel 10       & Kernel 20       \\
\hline
   1000 & 0.50 $\pm$ 0.01 & 0.49 $\pm$ 0.01 & 0.50 $\pm$ 0.01 & 0.50 $\pm$ 0.01 & 0.51 $\pm$ 0.01 & 0.48 $\pm$ 0.01 & 0.49 $\pm$ 0.01 & 0.48 $\pm$ 0.01 & 0.49 $\pm$ 0.01 \\
  10000 & 0.49 $\pm$ 0.01 & 0.50 $\pm$ 0.01 & 0.51 $\pm$ 0.01 & 0.51 $\pm$ 0.01 & 0.51 $\pm$ 0.01 & 0.50 $\pm$ 0.01 & 0.50 $\pm$ 0.01 & 0.50 $\pm$ 0.01 & 0.50 $\pm$ 0.01 \\
 100000 & 0.51 $\pm$ 0.01 & 0.50 $\pm$ 0.01 & 0.50 $\pm$ 0.01 & 0.50 $\pm$ 0.01 & 0.51 $\pm$ 0.01 & 0.50 $\pm$ 0.01 & 0.51 $\pm$ 0.01 & 0.50 $\pm$ 0.01 & 0.50 $\pm$ 0.01 \\
\hline
\end{tabular}
			\end{adjustbox}		
		\end{center}
	\end{table}

	\begin{table}[htb]
		\caption{Tow Moons: runtime in seconds; recall that GAN was trained on GPU while the SR methods were trained on a single CPU.}% Cal. error and NRMSE: the smaller, the better. R$ ^2 		
		\label{tab:Two_Moons_runtime}
		\begin{center}
			\begin{adjustbox}{max width=\textwidth}
				\begin{tabular}{rrrrrrrrrr}
\hline
        &   GAN &   Energy 3 &   Energy 5 &   Energy 10 &   Energy 20 &   Kernel 3 &   Kernel 5 &   Kernel 10 &   Kernel 20 \\
\hline
   1000 &  4799 &        578 &        690 &         759 &         896 &        585 &        613 &         651 &         852 \\
  10000 &  8163 &       1775 &       1917 &        2415 &        3228 &       1708 &       1883 &        2329 &        3267 \\
 100000 & 30232 &       9266 &       9388 &        9903 &       10805 &       9283 &       9479 &        9859 &       10902 \\
\hline
\end{tabular}
			\end{adjustbox}		
		\end{center}
	\end{table}
	
	\begin{table}[htb]
		\caption{Two Moons: epoch at which early stopping occurred; the max number of training epochs was 20000.}
		\label{tab:Two_Moons_es}
		\begin{center}
			\begin{adjustbox}{max width=\textwidth}
				\begin{tabular}{rrrrrrrrrr}
\hline
        &   GAN &   Energy 3 &   Energy 5 &   Energy 10 &   Energy 20 &   Kernel 3 &   Kernel 5 &   Kernel 10 &   Kernel 20 \\
\hline
   1000 & 20000 &      20000 &      20000 &       20000 &       20000 &      20000 &      20000 &       20000 &       20000 \\
  10000 & 20000 &      20000 &      20000 &       20000 &       20000 &      20000 &      20000 &       20000 &       20000 \\
 100000 & 20000 &      20000 &      20000 &       20000 &       20000 &      20000 &      20000 &       20000 &       20000 \\
\hline
\end{tabular}
			\end{adjustbox}		
		\end{center}
	\end{table}
	
		\FloatBarrier
	\subsection{Shallow Water Model}\label{app:res_shallow_water}
		\FloatBarrier

\begin{figure}
		\centering
		\includegraphics[width=\linewidth]{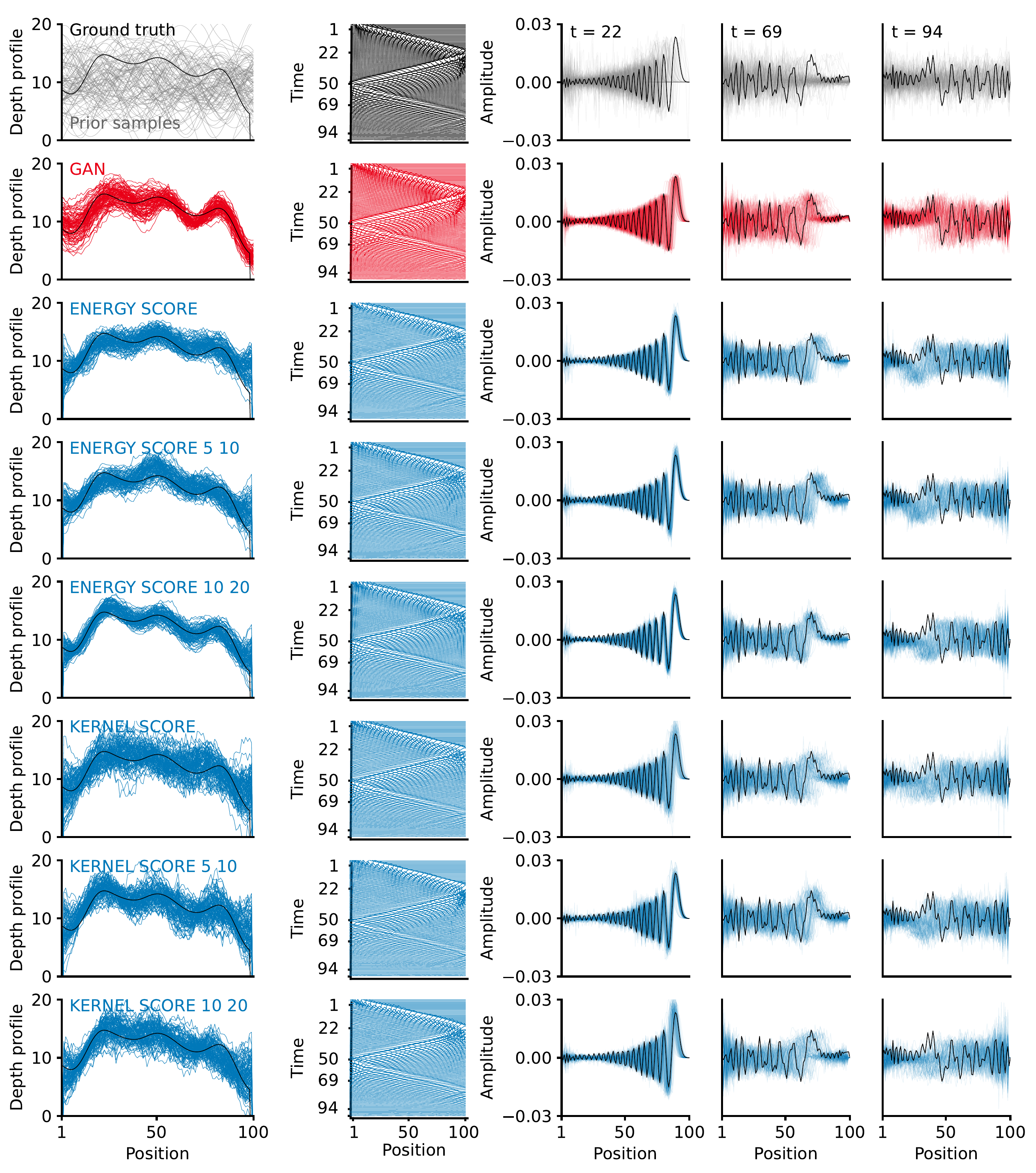}
		\caption{Shallow water model: inference results with all methods. See Figure~\ref{fig:shallow_water_model_best} for description of the different panels.}
		\label{fig:shallow_water_model_all}
	\end{figure}

\begin{figure}[ht]
		\centering
		\begin{subfigure}[t]{0.5\textwidth}
			\begin{center}
				\includegraphics[width=\columnwidth]{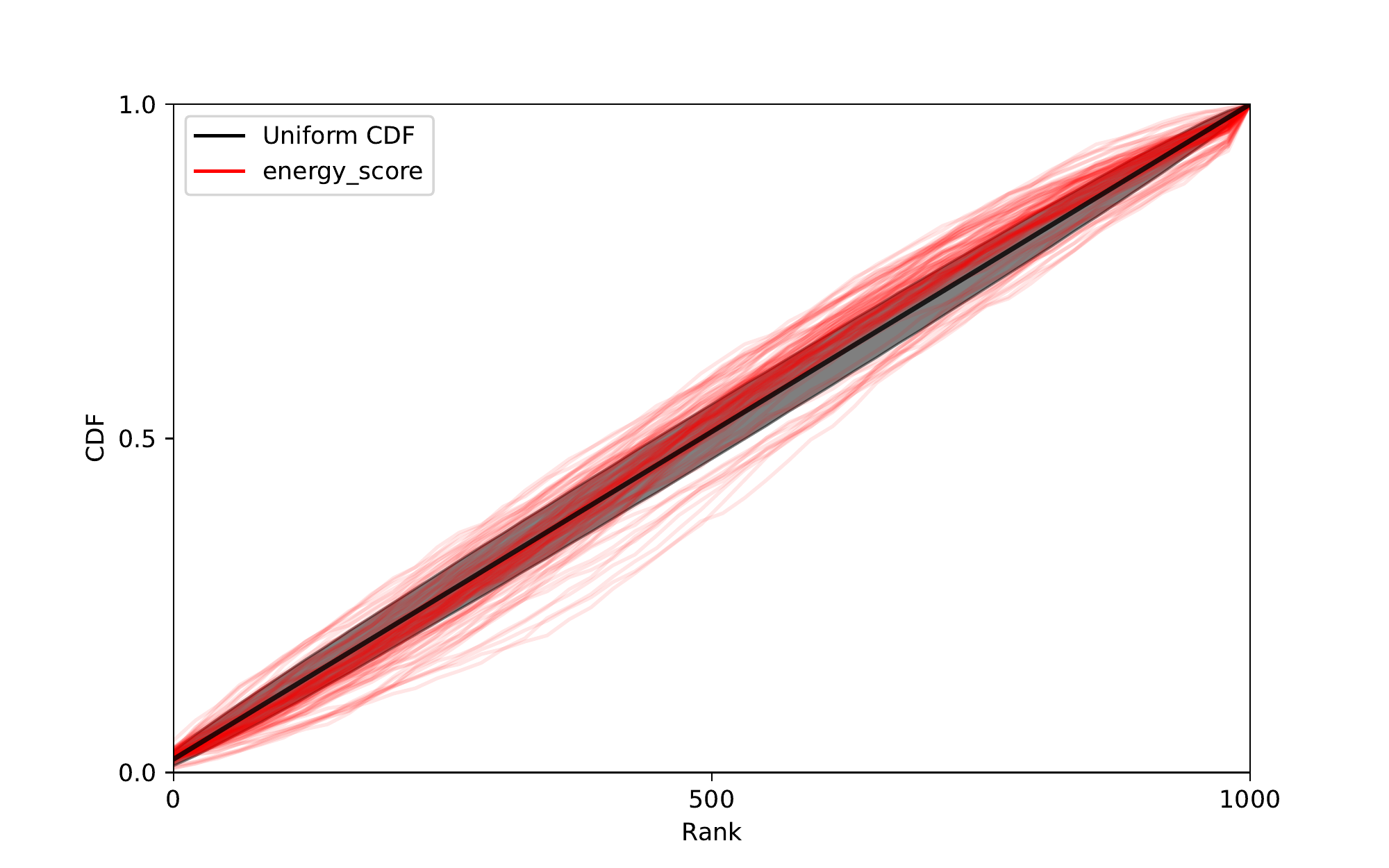} 
			\end{center}
			\caption{Energy Score}\label{}
		\end{subfigure}%
		\begin{subfigure}[t]{0.5\textwidth}
			\centering
			\includegraphics[width=\columnwidth]{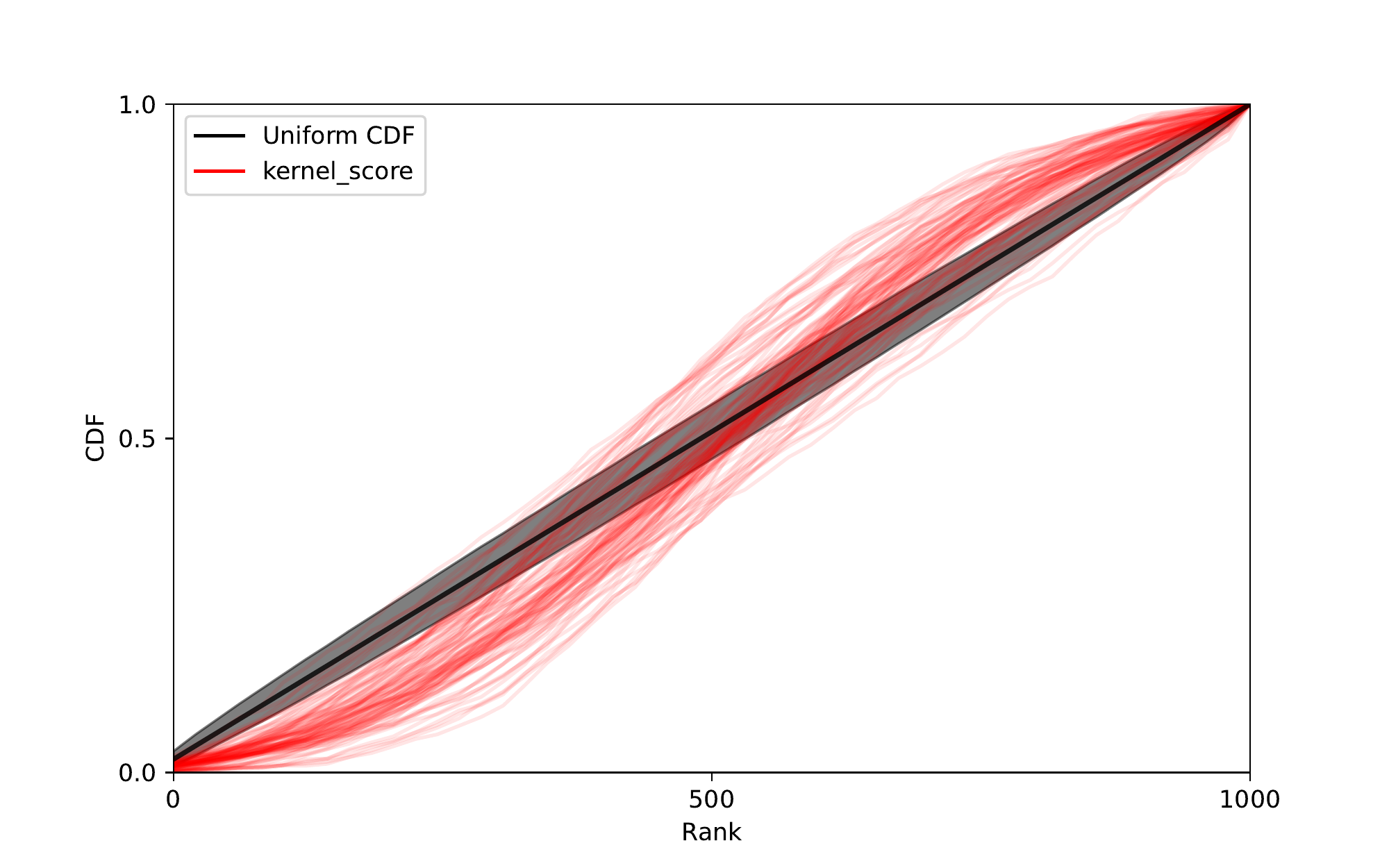}
			\caption{Kernel Score}\label{}
		\end{subfigure}\\
		\begin{subfigure}[t]{0.5\textwidth}
			\begin{center}
				\includegraphics[width=\columnwidth]{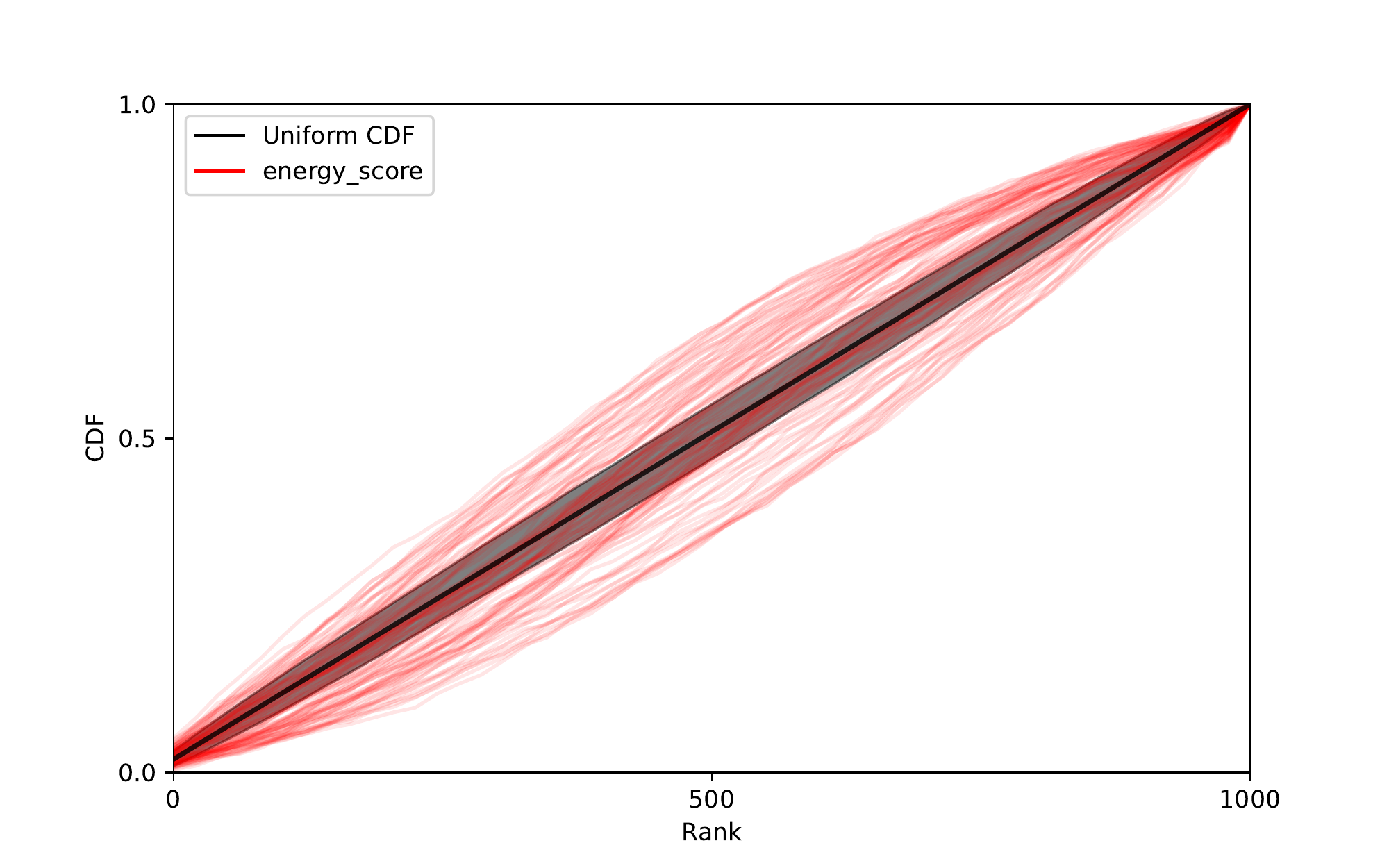} 
			\end{center}
			\caption{Energy Score patched 5, 10}\label{}
		\end{subfigure}%
		\begin{subfigure}[t]{0.5\textwidth}
			\centering
			\includegraphics[width=\columnwidth]{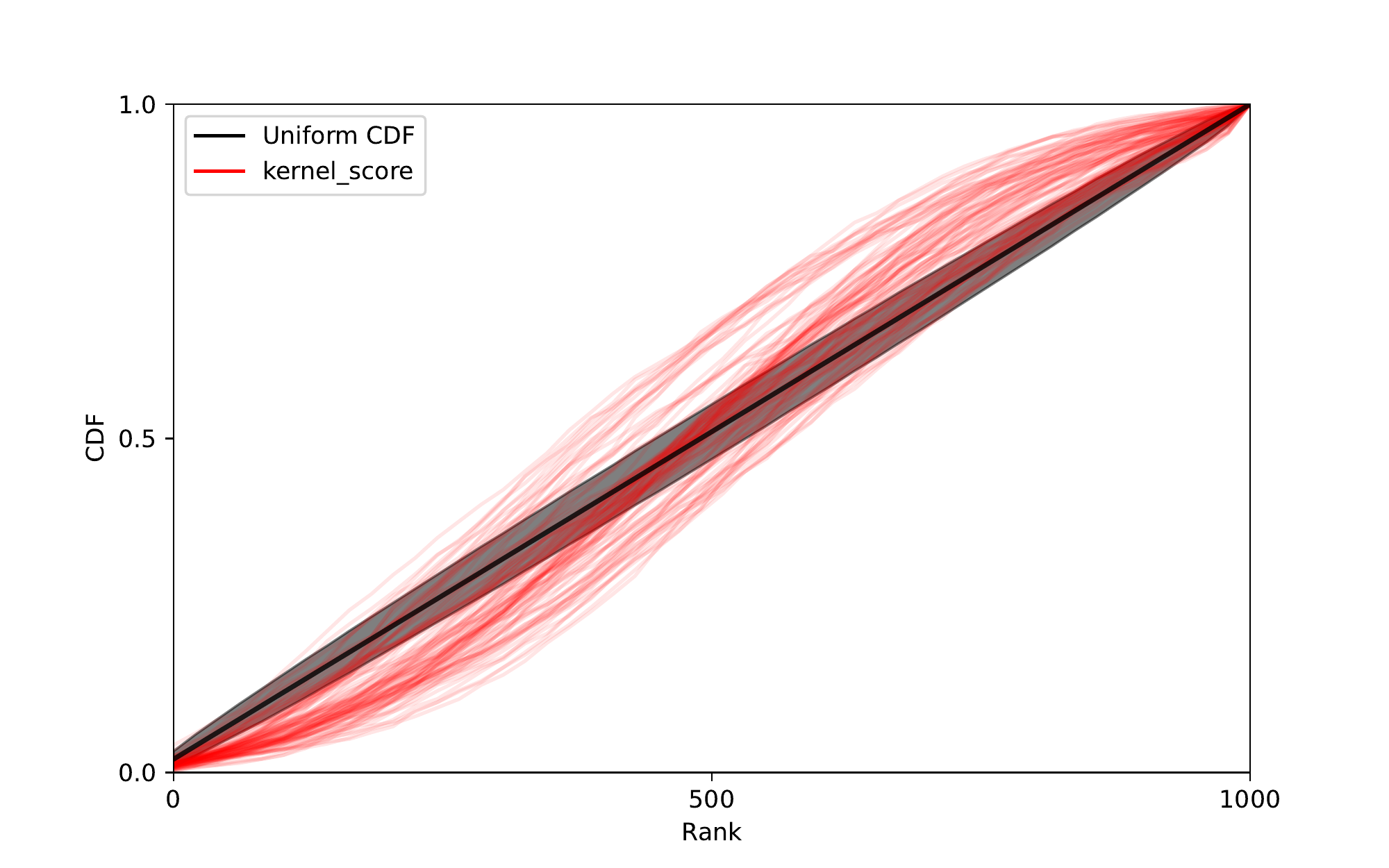} 
			\caption{Kernel Score patched 5, 10}\label{}
		\end{subfigure}\\
		\begin{subfigure}[t]{0.5\textwidth}
			\begin{center}
				\includegraphics[width=\columnwidth]{fig/shallow_water_sbc/energy_score_10_100k_patched_10_20_sbc_lines.pdf} 
			\end{center}
			\caption{Energy Score patched  10, 20}\label{}
		\end{subfigure}%
		\begin{subfigure}[t]{0.5\textwidth}
			\centering
			\includegraphics[width=\columnwidth]{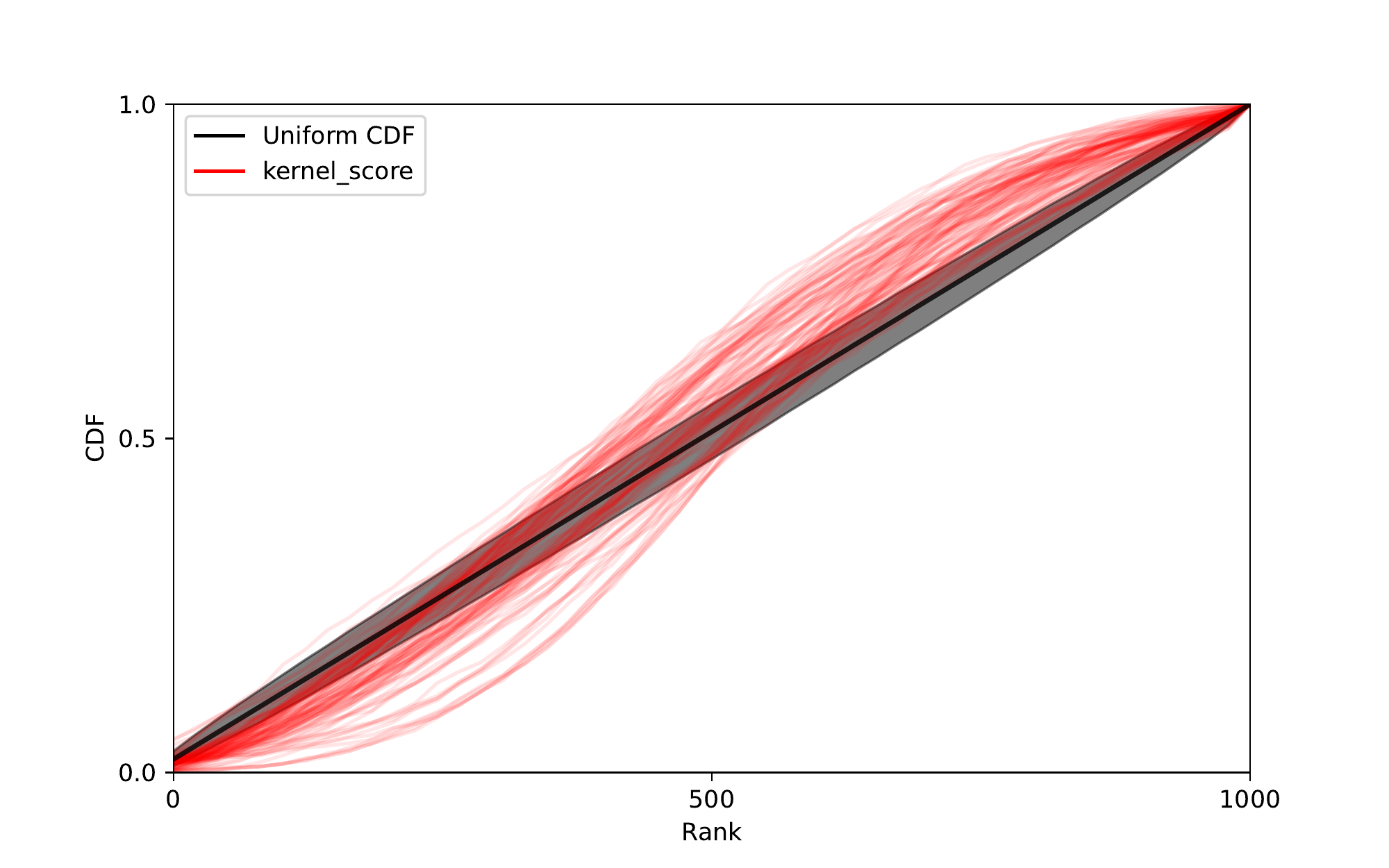} 
			\caption{Kernel Score patched 10, 20}\label{}
		\end{subfigure}\\
		\caption{Shallow Water model: Simulation Based Calibration for all SR methods. Each line corresponds to a single dimension of $ \ddtheta $ and represents the CDF of the rank of the true parameter value with respect to a set of posterior samples. A calibrated posterior implies uniform CDF (diagonal black line, with associated 99\% confidence region for that number of samples in gray).}
		\label{fig:SBC_shallow_water_all}
	\end{figure}

	In Figure~\ref{fig:shallow_water_model_all}, we show results, analogously to what done in Figure~\ref{fig:shallow_water_model_best}, for all methods. Table~\ref{tab:shallow_water_model_all} reports the different performance metrics, the runtime and the early stopping epoch for all methods. Finally, Figure~\ref{fig:SBC_shallow_water_all} reports Simulation-Based Calibration results for all SR methods.

	% all results for shallow water model

	\begin{table}[htb]
		\caption{Shallow Water model: performance metrics, runtime and early stopping epoch for all methods. We do not train GAN from scratch but rather relied on the trained network obtained in \cite{ramesh2022gatsbi}. The training time we report here corresponds to what is mentioned in the \cite{ramesh2022gatsbi}, which used two GPUs for training (with respect to a single one for the SR methods). For the same reason, we do not report the epoch at which GAN training was early stopped.}% Cal. error and NRMSE: the 	
		\label{tab:shallow_water_model_all}
		\begin{center}
			\begin{adjustbox}{max width=\textwidth}
				\begin{tabular}{llllrr}
\hline
                          & RMSE $ \downarrow $   & Cal. Err. $ \downarrow $   & R$^2$ $ \uparrow $   &   Runtime (sec) &   Early stopping epoch \\
\hline
 Energy               & 0.05 $\pm$ 0.01       & 0.03 $\pm$ 0.02            & 0.87 $\pm$ 0.05      &           51328 &                  10400 \\
 Energy patched 10 20 & 0.05 $\pm$ 0.01       & 0.03 $\pm$ 0.02            & 0.89 $\pm$ 0.05      &           60017 &                  12400 \\
 Energy patched 5 10  & 0.06 $\pm$ 0.01       & 0.03 $\pm$ 0.02            & 0.86 $\pm$ 0.06      &           49626 &                   9600 \\
 Kernel               & 0.06 $\pm$ 0.01       & 0.11 $\pm$ 0.05            & 0.84 $\pm$ 0.06      &           39608 &                   7800 \\
 Kernel patched 10 20 & 0.06 $\pm$ 0.01       & 0.09 $\pm$ 0.04            & 0.86 $\pm$ 0.06      &           47642 &                   9000 \\
 Kernel patched 5 10  & 0.06 $\pm$ 0.01       & 0.09 $\pm$ 0.04            & 0.86 $\pm$ 0.06      &           44590 &                   9200 \\
GAN & 0.07 $\pm$ 0.01   & 0.12 $\pm$ 0.09   & 0.78 $\pm$ 0.05 & $ \approx $345600 & - \\
\hline
\end{tabular}
			\end{adjustbox}		
		\end{center}
	\end{table}

	% SBC for all my methods: 

	% SBC lines for shallow water model; would need to compare with the GAN one as well. 

	\FloatBarrier
	\subsection{Camera model}\label{app:res_camera}
	\FloatBarrier

\begin{figure}
		\centering
		\includegraphics[width=\linewidth]{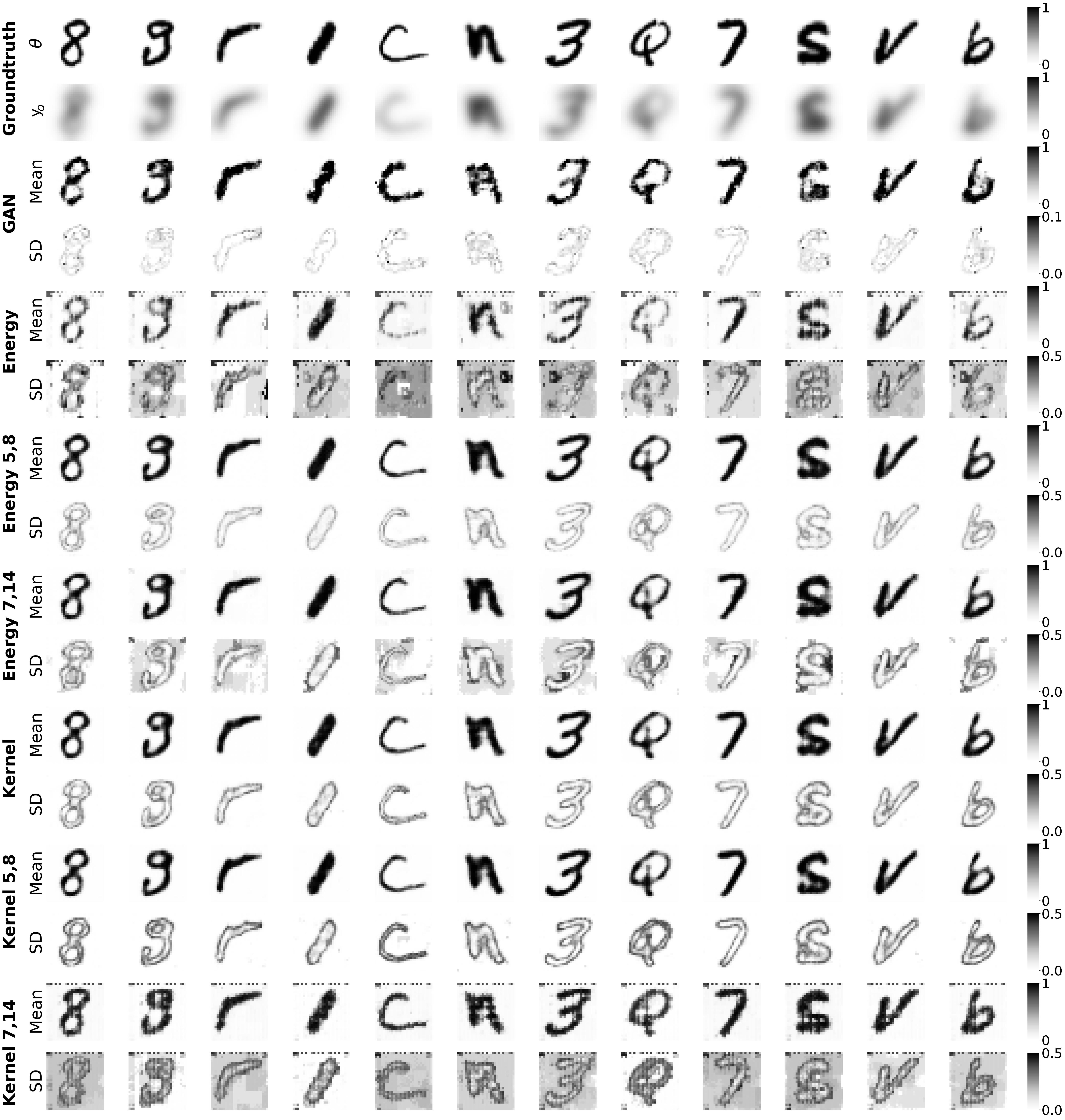}
		\caption{Noisy Camera model: ground truth and posterior inference with all methods, for a set of observations (each observation corresponds to a column). The first two rows represent the ground-truth values of $ \ddtheta $ and the corresponding observation $ \ddobs_o $. The remaining rows represent mean and Standard Deviation (SD) for all methods.}
		\label{fig:camera_model_all}
	\end{figure}

	In Figure~\ref{fig:camera_model_all}, we show results, analogously to what done in Figure~\ref{fig:camera_model_best}, for all methods. Table~\ref{tab:camera_model_all} reports the different performance metrics, the runtime and the early stopping epoch for all methods.

	% all results for camera model
	
	\begin{table}[htb]
	\caption{Noisy Camera model: performance metrics, runtime and early stopping epoch for all methods.}
				\label{tab:camera_model_all}
		\begin{center}
			\begin{adjustbox}{max width=\textwidth}
\begin{tabular}{llllrr}
	\hline
	& RMSE $ \downarrow $   & Cal. Err. $ \downarrow $   & R$^2$ $ \uparrow $    &   Runtime (sec) &   Early stopping epoch \\
	\hline
	GAN  &
	0.25 $\pm$ 0.19       & 0.50 $\pm$ 0.00            & -23.94 $\pm$ 366.08   &           45398 &                   3600 \\
	Energy              & 0.08 $\pm$ 0.05       & 0.36 $\pm$ 0.12            & -24.39 $\pm$ 450.13   &           24555 &                   4200 \\
	Energy patched 5 8  & 0.06 $\pm$ 0.05       & 0.36 $\pm$ 0.12            & -2.14 $\pm$ 55.86     &           22633 &                   4000 \\
	Energy patched 7 14 & 0.07 $\pm$ 0.05       & 0.37 $\pm$ 0.12            & -10.33 $\pm$ 227.38   &           24033 &                   3600 \\
	Kernel              & 0.06 $\pm$ 0.05       & 0.32 $\pm$ 0.15            & -7.22 $\pm$ 164.26    &           21862 &                   3200 \\
	Kernel patched 5 8  & 0.07 $\pm$ 0.05       & 0.36 $\pm$ 0.12            & -10.29 $\pm$ 222.12   &           22545 &                   3200 \\
	Kernel patched 7 14 & 0.10 $\pm$ 0.06       & 0.38 $\pm$ 0.11            & -144.56 $\pm$ 2952.80 &           20605 &                   3600 \\
	\hline
\end{tabular}			\end{adjustbox}		
		\end{center}
	\end{table}

\end{document}